\documentclass[prd,aps,superscriptaddress,10pt,nobibnotes,notitlepage,twocolumn,longbibliography,nofootinbib,preprintnumbers]{revtex4-1}

\usepackage[normalem]{ulem}
\usepackage{sidecap}
\usepackage[latin1]{inputenc}
\usepackage{graphicx}
\usepackage{float}
\usepackage{latexsym}
\usepackage{graphicx}
\usepackage{amssymb}
\usepackage{amsmath}
\usepackage{amsfonts}
\usepackage{ragged2e}

\usepackage[colorlinks=true,citecolor=blue,hyperfootnotes=false]{hyperref}

\usepackage{cool}
\usepackage{dcolumn}
\usepackage{textcomp}
\usepackage[dvipsnames]{xcolor}
\usepackage{xfrac}
\usepackage{slashed}
\usepackage{multirow}

\def\be{\begin{equation}}
\def\ee{\end{equation}}
\def\figs/B{B}
\def\bea{\begin{eqnarray}}
\def\eea{\end{eqnarray}}
\def\bg{\begin{eqnarray}}
\def\nd{\end{eqnarray}}
\def\sin{{\rm sin}}
\def\cos{{\rm cos}}

\def\log{{\rm log}}
\def\ln{{\rm log}}

\def\beq{\begin{equation}}
\def\eeq{\end{equation}}
\def\n{\nonumber \\}

\begin{document}

\preprint{MIT-CTP/5525} 

\title{Planck Constraints and Gravitational Wave Forecasts for \\Primordial Black Hole Dark Matter Seeded by Multifield Inflation
}

\author{Wenzer Qin}
\email{wenzerq@mit.edu}
\affiliation{Department of Physics, Massachusetts Institute of Technology, Cambridge, MA 02139, USA}

\author{Sarah R.~Geller}
\email{sgeller@mit.edu}
\affiliation{Department of Physics, Massachusetts Institute of Technology, Cambridge, MA 02139, USA}

\author{Shyam Balaji}
\email{sbalaji@lpthe.jussieu.fr}
\affiliation{Laboratoire de Physique Th\'{e}orique et Hautes Energies (LPTHE),
UMR 7589 CNRS \& Sorbonne Universit\'{e}, 4 Place Jussieu, F-75252, Paris, France}
\affiliation{Institut d'Astrophysique de Paris, UMR 7095 CNRS \& Sorbonne Universit\'{e}, 98 bis boulevard Arago, F-75014 Paris, France}

\author{Evan McDonough}
\email{e.mcdonough@uwinnipeg.ca}
\affiliation{Department of Physics, University of Winnipeg, Winnipeg MB, R3B 2E9, Canada}

\author{David I.~Kaiser}
\email{dikaiser@mit.edu}
\affiliation{Department of Physics, Massachusetts Institute of Technology, Cambridge, MA 02139, USA}

\begin{abstract}
We perform a Markov Chain Monte Carlo (MCMC) analysis of a simple yet generic multifield inflation model characterized by two scalar fields coupled to each other and nonminimally coupled to gravity, fit to {\it Planck} 2018 cosmic microwave background (CMB) data. In particular, model parameters are constrained by data on the amplitude of the primordial power spectrum of scalar curvature perturbations on CMB scales $A_s$, the spectral index $n_s$, and the ratio of power in tensor to scalar modes $r$, with a prior that the primordial power spectrum should also lead to primordial black hole (PBH) production sufficient to account for the observed dark matter (DM) abundance. We find that $n_s$ in particular controls the constraints on our model. Whereas previous studies of PBH formation from an ultra-slow-roll phase of inflation have highlighted the need for at least one model parameter to be highly fine-tuned, we identify a degeneracy direction in parameter space such that shifts by $\sim 10\%$ of one parameter can be compensated by comparable shifts in other parameters while preserving a close fit between model predictions and observations. Furthermore, we find this allowed parameter region produces observable gravitational wave (GW) signals in the frequency ranges to which upcoming experiments are projected to be sensitive, including Advanced LIGO and Virgo, the Einstein Telescope (ET), Cosmic Explorer (CE), DECIGO, and LISA.
\end{abstract}

\date{\today}

\maketitle

\section{Introduction}

Perhaps the most prominent and enduring puzzle in modern cosmology is the unknown nature of dark matter (DM), the predominant form of matter in the universe that is vital to structure formation and galactic stability. 
Primordial black holes (PBHs) \cite{Zeldovich:1967lct,Hawking:1971ei,Carr:1974nx,Meszaros:1974tb,Carr:1975qj,Khlopov:1985jw,Niemeyer:1999ak} have been posited to make up anywhere from less than a percent to the entirety of the DM abundance. 

A leading candidate for a mechanism to seed PBH formation in the early universe is an amplification of the primordial spectrum of curvature perturbations. The typical mass with which PBHs form scales with the Hubble mass---the mass-energy contained within a Hubble sphere at the time of PBH formation---and hence PBHs could be a plausible explanation for a diverse range of phenomena. For example, a population of PBHs with $M_{\rm pbh} \sim {\cal O} (10) \, M_\odot$ could be relevant for various binary black hole merger events reported by LIGO-Virgo, whereas PBHs with $M_{\rm pbh} \sim {\cal O} (10^5) \, M_\odot$ could have served as seeds for the supermassive black holes at the centers of galaxies. On the other hand, if PBHs are to account for the entire observed DM abundance, then various theoretical and observational constraints limit the mass range for PBHs to a much lighter regime, with $10^{-16} \, M_\odot \leq M_{\rm pbh} \leq 10^{-11} \, M_\odot$. See Refs.~\cite{Khlopov:2008qy,Sasaki:2018dmp,Carr:2020gox,Carr:2020xqk,Green:2020jor,Escriva:2021aeh,Villanueva-Domingo:2021spv,Escriva:2022duf,Ozsoy:2023ryl} for recent reviews of PBHs and their current observational status.  

Given the success of cosmic inflation as a model of the early universe, it is natural to turn to inflation as a mechanism to amplify the primordial power spectrum. The vast majority of work on PBHs formed after a phase of cosmic inflation has focused on single-field inflationary models. (See, e.g., the recent review article in ~Ref.~\cite{Ozsoy:2023ryl} and references therein.) These constructions typically include {\it ad hoc} features in the inflationary potential $V(\phi)$, which could be better motivated in the context of additional fields, such as in the case of a step-feature model \cite{Inomata:2021tpx,Inomata:2021uqj} or renormalization-group flow generated by interactions among several distinct fields 
\cite{Ezquiaga:2017fvi}. 
Moreover, in such multifield models, it is generally not the case that a single scalar field can be isolated as the sole dynamical degree of freedom, with a mass hierarchy both above and below the scale of interest. Indeed, even the Standard Model of particle physics features four real scalars, namely the real components of the Higgs doublet, each of which remain in the spectrum at high energies in renormalizable gauges \cite{Mooij:2011fi,Greenwood:2012aj,Kawai:2015ryj}. Finally, whereas single-field inflation models for PBH formation often appear finely-tuned and therefore `unnatural,' there remains the possibility that such models may become natural when viewed from the perspective of a multifield model.

These considerations motivate the study of PBH formation directly in the context of multifield inflation. In a recent paper by the authors (Ref.~\cite{Geller:2022nkr}), we demonstrated that PBHs relevant for DM will form from a realistic class of multifield inflationary models built from well-motivated high-energy ingredients. In particular, we considered inflationary models that incorporate multiple interacting scalar fields, each with a nonminimal gravitational coupling. Such models consist of generic mass-dimension-4 operators in an effective field theory (EFT) expansion of the action at inflationary energy scales, and (as discussed in Ref.~\cite{Geller:2022nkr}) have a consistent ultraviolet completion in the context of supergravity. For various choices of model parameters, the inflationary dynamics in such models can include a phase of ultra-slow-roll evolution \cite{Garcia-Bellido:2017mdw,Ezquiaga:2017fvi,Germani:2017bcs,Kannike:2017bxn,Motohashi:2017kbs,Di:2017ndc,Ballesteros:2017fsr,Pattison:2017mbe,Passaglia:2018ixg,Biagetti:2018pjj,Byrnes:2018txb,Carrilho:2019oqg,Figueroa:2020jkf,Inomata:2021tpx,Inomata:2021uqj,Pattison:2021oen,Balaji:2022rsy,Balaji:2022zur,Kawai:2022emp,Karam:2022nym,Pi:2022ysn}, which can yield PBHs with $M_{\rm pbh}$ within the appropriate range to account for the entire DM abundance, while also matching high-precision measurements of the primordial perturbation spectrum on length-scales relevant for the cosmic microwave background radiation (CMB). In Ref.~\cite{Geller:2022nkr}, we demonstrated that eight distinct observational constraints---relating to both PBH and CMB observables---could be matched by adjusting only six free parameters in these models.

In this paper, we use a Markov Chain Monte Carlo (MCMC) analysis to systematically identify the regions of parameter space for the family of models constructed in Ref.~\cite{Geller:2022nkr} that can produce relevant populations of PBHs for DM while continuing to match multiple observables related to the CMB. 
Our two-dimensional marginalized posterior distributions for pairs of parameters demonstrate that this general class of models can yield predictions for observables near the CMB pivot scale $k_* = 0.05 \, {\rm Mpc}^{-1}$ in close agreement with the latest observations \cite{Planck:2018jri,Planck:2018vyg,Planck:2019kim,BICEP:2021xfz}---including the amplitude of the primordial power spectrum $A_s (k_*)$, the spectral tilt $n_s (k_*)$, and the ratio of power in tensor to scalar modes $r (k_*)$---while also producing PBHs with masses $M_{\rm pbh}$ within the range for which they could account for the entire DM abundance. Such regions of parameter space also yield predictions for related observables that are easily compatible with the latest observational bounds, such as the running of the spectral index $\alpha_s (k_*)$, the primordial isocurvature fraction $\beta_{\rm iso} (k_*)$, and local primordial non-Gaussianity $f_{\rm NL}$. 

In Ref.~\cite{Geller:2022nkr}, much as in previous studies of PBH formation from ultra-slow-roll evolution \cite{Garcia-Bellido:2017mdw,Ezquiaga:2017fvi,Germani:2017bcs,Kannike:2017bxn,Motohashi:2017kbs,Di:2017ndc,Ballesteros:2017fsr,Pattison:2017mbe,Passaglia:2018ixg,Biagetti:2018pjj,Byrnes:2018txb,Carrilho:2019oqg,Figueroa:2020jkf,Inomata:2021tpx,Inomata:2021uqj,Pattison:2021oen}, we found that one model parameter needed to be highly fine-tuned in order to match predictions for all eight relevant observables. As we will discuss here, with the aid of the MCMC we identify a degeneracy direction in parameter space such that shifts up to $\sim 10\%$ of any particular model parameter can be compensated by comparable shifts among the other parameters while preserving a close fit with observations. The overall tuning of each parameter required to match all the observables of interest, as measured by the posterior distributions for various {\it ratios} of parameters, is at the percent level, driven largely by the sub-percent-level accuracy to which the spectral index $n_s$ has been measured.

In addition to studying predictions from these models for CMB observables and PBH formation, we also analyze predictions for the amplification of primordial gravitational waves (GWs). GWs provide a tantalizing means to test the physics of the early universe. (For recent reviews, see Refs.~\cite{Guzzetti:2016mkm,Caprini:2018mtu,Domenech:2021ztg}.) In the context of PBH formation, the amplified spectrum of scalar curvature perturbations---necessary to induce gravitational collapse into PBHs---will source tensor modes beyond linear order in perturbation theory. Therefore PBH formation should be accompanied by a contribution to a stochastic GW background (SGWB) with a particular spectral shape \cite{Saito:2008jc,Saito:2009jt,Assadullahi:2009jc,Bugaev:2009zh,Bugaev:2010bb,Inomata:2018epa}. 

Whereas the primordial GW spectrum is tightly constrained on scales near the CMB pivot scale $k_*$, it is mostly unconstrained on the much shorter length-scales relevant for PBH formation. We calculate the expected contribution to the SGWB from PBH formation in our models, and find that the signal overlaps significantly with the (projected) integrated sensitivity curves for several next-generation detectors, including Advanced LIGO-Virgo (LIGO A+) \cite{KAGRA:2021kbb}, LISA \cite{Barausse:2020rsu}, the Einstein Telescope (ET) \cite{Maggiore:2019uih}, Cosmic Explorer (CE)~\cite{Reitze:2019iox}, and DECIGO \cite{Yuan:2021qgz,Kawamura:2020pcg}. These results suggest the exciting possibility that the production of DM in the form of PBHs from multifield models could soon be testable. 

This paper is organized as follows. In \S~\ref{sec:model}, we introduce the multifield inflationary model and discuss dynamics during inflation. In \S~\ref{sec:observables}, we discuss relevant physical observables predicted by the model. 
In \S~\ref{sec:constraints}, we discuss how we constrain the allowed model parameter space.
In \S~\ref{sec:results} we present our results and finally we conclude with further discussion in \S~\ref{sec:conclusion}. We discuss the effects of a phase of ultra-slow-roll evolution on the power spectrum of scalar curvature perturbations in Appendix \ref{app:USR}, and present additional details on the calculation of the induced GW spectrum in Appendix \ref{app:tensor_spectrum}.

\section{Multifield inflation model}
\label{sec:model}

Many types of single-field inflationary models yield predictions for CMB observables that are consistent with current observations. (See, e.g., Refs.~\cite{Bassett:2005xm,Martin:2013tda,Guth:2013sya,Planck:2018jri}.) 
Single-field models can also produce populations of PBHs relevant for DM \cite{Carr:2020gox,Carr:2020xqk,Green:2020jor,Escriva:2022duf,Escriva:2022duf,Ozsoy:2023ryl,Kawai:2021edk,Ahmed:2021ucx}. On the other hand, the Standard Model of particle physics includes multiple scalar degrees of freedom (at high energies in renormalizable gauges), and extensions beyond the Standard Model generically include many more \cite{Mooij:2011fi,Greenwood:2012aj,Lyth:1998xn,Mazumdar:2010sa,Balaji:2022dbi}. We therefore focus on multifield models.

Likewise, nonminimal couplings in the effective action of the form $\xi \phi^2 R$, where $R$ is the spacetime Ricci scalar and $\xi$ is a dimensionless constant, are required for renormalization and are induced at one-loop even if the couplings $\xi$ vanish at tree level \cite{Callan:1970ze,Bunch:1980br,Bunch:1980bs,Birrell:1982ix,Parker:2009uva,Markkanen:2013nwa,Kaiser:2015usz}. The couplings $\xi$ generically increase with energy scale under renormalization-group flow with no UV fixed point \cite{Odintsov:1990mt,Buchbinder:1992rb}, and hence they can be large ($\vert \xi \vert \gg 1$) at inflationary energy scales. Hence in this work we focus on models with multiple interacting scalar fields, each with a nonminimal coupling to gravity.

Such models are natural generalizations of the multifield models studied in Refs.~\cite{Kaiser:2012ak,Kaiser:2013sna,Schutz:2013fua,Kaiser:2015usz}, and are closely related to well-known models such as Higgs inflation \cite{Bezrukov:2007ep,Greenwood:2012aj,Kawai:2014gqa,Kawai:2015ryj} and $\alpha$-attractors \cite{Kallosh:2013hoa,Kallosh:2013maa,Galante:2014ifa}. In addition to providing an excellent fit to CMB observables, such models also feature efficient reheating \cite{Bezrukov:2008ut,Garcia-Bellido:2008ycs,Child:2013ria,DeCross:2015uza,DeCross:2016cbs,DeCross:2016fdz,Figueroa:2016dsc,Repond:2016sol,Ema:2016dny,Sfakianakis:2018lzf,Rubio:2019ypq,Nguyen:2019kbm,vandeVis:2020qcp,Iarygina:2020dwe,Ema:2021xhq}.

We restrict attention to $(3+1)$ spacetime dimensions and use metric signature $(-,+,+,+)$. We also adopt natural units $\hbar = c = k_B = 1$, within which the reduced Planck mass is $M_{\rm pl} \equiv 1 / \sqrt{ 8 \pi G} = 2.43 \times 10^{18} \, {\rm GeV}$.

\subsection{Action and Equations of Motion}

In the Jordan frame, the action for ${\cal N}$ interacting scalar fields $\phi^I (x^\mu)$ with $I = 1, 2, ... , {\cal N}$, each with a nonminimal coupling to the spacetime Ricci scalar, is given by
\beq
\tilde{S} = \int d^4 x \sqrt{-\tilde{g}} \left[ f (\phi^I ) \tilde{R}  - \frac{1}{2} \delta_{IJ} \tilde{g}^{\mu\nu}  \partial_\mu \phi^I \partial_\nu \phi^J - \tilde{V} (\phi^I) \right] ,
\label{eqn:SJ}
\eeq
where $f (\phi^I )$ represents the nonminimal couplings, and tildes denote quantities in the Jordan frame.
Upon performing the conformal transformation
\beq
\tilde{g}_{\mu\nu} (x) \rightarrow g_{\mu\nu} (x) = \frac{2}{M_{\rm pl}^2} f (\phi^I (x) ) \, \tilde{g}_{\mu\nu} (x),
\label{eqn:Omega}
\eeq
we arrive at the Einstein-frame action~\cite{Kaiser:2010ps}
\beq
S = \int d^4 x \sqrt{-g} \left[ \frac{ M_{\rm pl}^2}{2} R - \frac{1}{2} {\cal G}_{IJ} g^{\mu\nu} \partial_\mu \phi^I \partial_\nu \phi^J - V (\phi^I) \right] .
\label{eqn:SE}
\eeq
The potential in the Einstein frame, $V (\phi^I)$, is related to the potential in the Jordan frame, $\tilde{V} (\phi^I$), by
\beq
V (\phi^I) = \frac{ M_{\rm pl}^4}{4 f^2 (\phi^I) } \tilde{V} (\phi^I ) .
\label{eqn:VEconformal}
\eeq
In addition, the nonminimal couplings induce a nontrivial curvature of the field-space manifold in the Einstein frame, with metric given by 
\beq
{\cal G}_{IJ} (\phi^K) = \frac{M_{\rm pl}^2}{2 f (\phi^K)} \left[ \delta_{IJ} + \frac{ 3}{f (\phi^K)} f_{, I} f_{, J} \right] ,
\label{eqn:GIJgeneral}
\eeq
where $f_{, I} \equiv \partial f / \partial \phi^I$. If more than one field has a nonminimal coupling to $\tilde{R}$, then one cannot canonically normalize all of the fields $\phi^I$ while also retaining the simple Einstein-Hilbert term in the action \cite{Kaiser:2010ps}.

To study the dynamics of background-order quantities and linearized fluctuations in such models, we adopt the gauge-invariant multifield formalism of Refs.~\cite{Sasaki:1995aw,Langlois:2008mn,Peterson:2010mv,Peterson:2010np,Gong:2011cd,Gong:2011uw,Kaiser:2012ak,DeCross:2015uza,Gong:2016qmq}, and consider perturbations around a spatially flat Friedmann-Lema\^{i}tre-Robertson-Walker (FLRW) line element. Each scalar field may be decomposed into its vacuum expectation value and a spatially varying fluctuation,
\begin{equation}
    \phi^I (x^\mu) = \varphi^I (t) + \delta \phi^I (x^\mu).
    \label{phivarphi}
\end{equation}
The magnitude of the velocity of the background fields can be written as
\begin{equation}
    \vert \dot{\varphi}^I \vert \equiv \dot{\sigma} = \sqrt{ {\cal G}_{IJ} \, \dot{\varphi}^I \dot{\varphi}^J } ,
    \label{dotsigmadef}
\end{equation}
so the unit vector in the direction of the fields' velocity is given by
\begin{equation}
    \hat{\sigma}^I \equiv \frac{ \dot{\varphi}^I } {\dot{\sigma}} .
    \label{hatsigma}
\end{equation}
The covariant turn-rate vector is defined as
\beq
\omega^I \equiv {\cal D}_t \hat{\sigma}^I ,
\label{omegaturnratedef}
\eeq
where ${\cal D}_t A^I \equiv \dot{\varphi}^J {\cal D}_J A^I$ for any field-space vector $A^I$, and the covariant derivative ${\cal D}_J$ employs the usual Levi-Civita connection associated with the field-space metric ${\cal G}_{IJ}$.

In terms of these quantities, the equations of motion for the background fields in the Einstein frame are
\beq
\begin{split}
    \ddot{\sigma} &+ 3 H \dot{\sigma} + V_{, \sigma} = 0 , \\
    H^2 &= \frac{1}{ 3 M_{\rm pl}^2} \left[ \frac{1}{2} \dot{\sigma}^2 + V \right] , \\
    \dot{H} &= - \frac{1}{ 2 M_{\rm pl}^2} \dot{\sigma}^2 ,
\end{split}
    \label{eombackground}
\eeq
where $H \equiv \dot{a} / a$ and
\begin{equation}
    V_{, \sigma} \equiv \hat{\sigma}^I V_{, I} .
    \label{Vsigma}
\end{equation}
Given the evolution of $\dot{\sigma} (t)$ and $H (t)$, we can then calculate the slow-roll parameters,
\beq
\begin{split}
    \epsilon &\equiv - \frac{ \dot{H}}{H^2} = \frac{ 1}{2 M_{\rm pl}^2} \frac{ \dot{\sigma}^2}{H^2} , \\
    \eta &\equiv 2 \epsilon - \frac{ \dot{\epsilon}}{2 H \epsilon}.
\end{split}
    \label{eqn:slowroll}
\eeq

Lastly, we may calculate the dimensionless power spectrum for the gauge-invariant scalar curvature perturbations ${\cal R}$, defined as
\beq
{\cal P}_{\cal R} (k) \equiv \frac{ k^3}{2 \pi^2} \vert {\cal R}_k (t_{\rm end} ) \vert^2 ,
\label{PRdefgeneral}
\eeq
where $t_{\rm end}$ indicates the end of inflation. As discussed in Ref.~\cite{Geller:2022nkr}, within the family of models we are considering, the fields generically evolve within a local minimum or ``valley" of the potential in the Einstein frame, and therefore the isocurvature modes remain heavy throughout the duration of inflation, $\mu_s^2 \gg H^2$. Likewise, the covariant turn rate remains small, $\vert \omega^I \vert \ll H$. (See also Refs.~\cite{Kaiser:2012ak,Kaiser:2013sna,Schutz:2013fua,Kaiser:2015usz,DeCross:2015uza}.) Under these conditions, when the background fields $\varphi^I (t)$ undergo ordinary slow-roll evolution with $\epsilon, \vert \eta \vert \ll 1$, the power spectrum assumes the form~\cite{Gordon:2000hv,Wands:2002bn,Bassett:2005xm,Kaiser:2012ak,Geller:2022nkr}
\begin{equation}
    {\cal P}_{\cal R}^{\rm SR} (k) = \frac{ H^2 (t_k) }{8 \pi^2 M_{\rm pl}^2 \epsilon (t_k) } \left( \frac{ k}{a (t_k) H (t_k)} \right)^{3 - 2 \nu_{\rm SR}} \big[ 1 + {\cal O} (\epsilon) \big] ,
    \label{eqn:PRHepsilon}
\end{equation}
with $\nu_{\rm SR} = \frac{3}{2} + 3 \epsilon - \eta$. As discussed in Appendix \ref{app:USR}, during ordinary slow-roll the modes ${\cal R}_k (t)$ remain frozen after crossing outside the Hubble radius, $\dot{\cal R}_k \simeq 0$ for $k \ll a H$, so one may evaluate ${\cal P}_{\cal R}^{\rm SR} (k)$ for ${\cal R}_k (t_k) \simeq {\cal R}_k (t_{\rm end})$, where $t_k$ is the time when a mode of comoving wavenumber $k$ first crossed outside the Hubble radius during inflation:
\begin{align}
    k = a(t_k) H (t_k).\label{crossout}
\end{align}

Inflationary dynamics that yield a brief phase of ultra-slow-roll evolution, during which $\epsilon (t_{\rm usr}) \ll 1$, will generate a spike in the power spectrum ${\cal P}_{\cal R} (k)$ on corresponding wavenumbers $k_{\rm usr}$. Such large-amplitude perturbations, in turn, can produce PBHs upon re-entering the Hubble radius after the end of inflation \cite{Garcia-Bellido:2017mdw,Ezquiaga:2017fvi,Germani:2017bcs,Kannike:2017bxn,Motohashi:2017kbs,Di:2017ndc,Ballesteros:2017fsr,Pattison:2017mbe,Passaglia:2018ixg,Biagetti:2018pjj,Byrnes:2018txb,Carrilho:2019oqg,Figueroa:2020jkf,Inomata:2021tpx,Inomata:2021uqj,Pattison:2021oen}. The main effect from ultra-slow-roll on the amplitude of the power spectrum is captured by the usual slow-roll expression in Eq.~(\ref{eqn:PRHepsilon}), given the relationship ${\cal P}_{\cal R}^{\rm SR} (k) \propto 1 / \epsilon$. Additional growth in ${\cal P}_{\cal R} (k)$ for certain wavenumbers $k$, beyond that represented by ${\cal P}_{\cal R}^{\rm SR} (k)$, can also occur during ultra-slow-roll. As discussed in Section \ref{sec:data}, we have performed about 2 million simulations of the dynamics of this family of models across a broad region of parameter space. In order for this to be computationally tractable, we used the expression of Eq.~(\ref{eqn:PRHepsilon}) in our Markov Chain Monte Carlo analysis, which depends only on background-order quantities, and hence can be evaluated for a given point in parameter space very efficiently. In Appendix \ref{app:USR}, we bound the discrepancy in ${\cal P}_{\cal R} (k)$ that can arise from the ultra-slow-roll phase in this family of models within the relevant regions of parameter space. 

For discussion of possible effects from loop corrections on the power spectrum, compare Refs.~\cite{Cheng:2021lif,Kristiano_yokoyama_1,Kristiano_yokoyama_2,sayantan_1,sayantan_2,Cheng:2023ikq} with Refs.~\cite{Senatore_2010,Senatore:2012nq,Pimentel:2012tw,Senatore:2012ya,Ando:2020fjm,RiottoPBH,Firouzjahi:2023aum,Motohashi:2023syh,Firouzjahi:2023ahg,Franciolini:2023lgy,Tasinato:2023ukp}.

\subsection{Two-Field Model}

Our aim is to include a {\it generic} set of mass-dimension-4 operators in the effective action for a two-field model. Even with only two fields, such models include a large number of free parameters. To help organize the couplings for each possible term, we adopt a supersymmetric framework, as in Ref.~\cite{Geller:2022nkr}. For the energy scales of interest, with $H \lesssim 10^{-5} \, M_{\rm pl}$ during inflation, we may consider the global supersymmetry limit of supergravity. 

We begin with a superpotential $\tilde{W} (\Phi)$ that includes only bilinear and trilinear couplings among two chiral superfields $\Phi^I$, where a tilde denotes quantities in the Jordan frame. Utilizing a shift-symmetric K\"{a}hler potential $\tilde{K} (\Phi, \bar{\Phi})$, as in countless supergravity inflation scenarios (see, e.g., Refs.~\cite{McDonough:2016der,Roest:2015qya}), one may easily construct models wherein the imaginary components of the scalar fields remain heavy during inflation, and hence effectively decouple. Upon relating the Jordan-frame potential $\tilde{V} (\phi, \chi)$ for the real-valued scalar fields $\phi$ and $\chi$ to $\sum_I \vert \partial \tilde{W} / \partial \Phi^I \vert^2$ in the usual way, this supersymmetric framework yields an expression for $\tilde{V} (\phi, \chi)$ of the form~\cite{Geller:2022nkr}
\begin{eqnarray}
    \tilde{V}(\phi,\chi) =&& 4 b_1^2 \mu ^2 \phi ^2+12 b_1 c_1 \mu  \phi ^3+8 b_1 c_2 \mu  \chi  \phi ^2 \nonumber \\
    && +4 b_1 c_3 \mu  \chi ^2 \phi +4 b_2^2 \mu ^2 \chi ^2+4 b_2 c_2 \mu  \chi  \phi ^2 \nonumber \\
    && +8 b_2 c_3 \mu  \chi ^2 \phi +12 b_2 c_4 \mu  \chi ^3+9 c_1^2 \phi ^4+12 c_1 c_2 \chi  \phi ^3 \nonumber \\
    && +6 c_1 c_3 \chi ^2 \phi ^2+4 c_2^2 \chi ^2 \phi ^2+c_2^2 \phi ^4+4 c_2 c_3 \chi  \phi ^3 \nonumber \\
    && +4 c_2 c_3 \chi ^3 \phi +6 c_2 c_4 \chi ^2 \phi ^2+c_3^2 \chi ^4+4 c_3^2 \chi ^2 \phi ^2 \nonumber \\
    && +12 c_3 c_4 \chi ^3 \phi +9 c_4^2 \chi ^4 ,
\end{eqnarray}
where $\mu$ has dimensions of mass and the parameters $\{ b_i, c_j\}$ are dimensionless couplings.
See Appendix B of Ref.~\cite{Geller:2022nkr} for further details.  

It is convenient to study the dynamics of our two-field model by adopting polar coordinates for the field space,
\beq
\phi (t) = r (t) \, \cos \theta (t) , \>\> \chi (t) = r (t) \, \sin \theta (t) ,
\label{eqn:phichirtheta}
\eeq
with $r \geq 0$ and $0 \leq \theta \leq 2 \pi$. Then the Jordan-frame potential can be written as
\beq
\tilde{V} (r, \theta) = {\cal B} (\theta) \mu^2 r^2 + {\cal C} (\theta) \mu r^3 + {\cal D} (\theta) r^4,
\label{eqn:Vtildertheta}
\eeq
where
\beq
\begin{split}
    {\cal B} (\theta) &\equiv 4 b_1^2 \cos^2 \theta + 4 b_2^2 \sin^2 \theta , \\
    {\cal C} (\theta) &\equiv 12 b_1 c_1 \cos^3 \theta + 4  (2b_1 + b_2) c_2 \cos^2 \theta \sin \theta \\
    &\quad + 4 (b_1 + 2 b_2) c_3 \cos \theta \sin^2 \theta + 12 b_2 c_4 \sin^3 \theta , \\
    {\cal D} (\theta) &\equiv (9 c_1^2 + c_2^2) \cos^4 \theta + 4 c_2 (3 c_1 + c_3) \cos^3 \theta \sin\theta \\
    &\quad + (4 c_2^2 + 6 c_1 c_3 + 6 c_2 c_4 + 4 c_3^2) \cos^2 \theta \sin^2\theta \\
    &\quad+ 4 c_3 (c_2 + 3 c_4) \cos \theta \sin^3 \theta + (9 c_4^2 + c_3^2) \sin^4 \theta .
\end{split}
\label{eqn:BCDdef}
\eeq
In addition to the couplings in the potential $\tilde{V}$, the fields also have nonminimal couplings to gravity,
\beq
\begin{split}
f (\phi, \chi) &= \frac{1}{2} \left[ M_{\rm pl}^2 + \xi_\phi \phi^2 + \xi_\chi \chi^2 \right] \\
&= \frac{1}{2} \left[ M_{\rm pl}^2 + r^2 \left( \xi_\phi \cos^2 \theta + \xi_\chi \sin^2 \theta \right) \right] ,
\end{split}
\label{eqn:frtheta}
\eeq
which are generated by the scalar fields' self-interactions in a curved spacetime~\cite{Callan:1970ze,Bunch:1980br,Bunch:1980bs,Birrell:1982ix,Odintsov:1990mt,Buchbinder:1992rb,Faraoni:2000gx,Parker:2009uva,Markkanen:2013nwa,Kaiser:2015usz}. Upon performing the conformal transformation as in Eq.~(\ref{eqn:Omega}), the potential in the Einstein frame takes the form
\beq
V (r, \theta) = \frac{ M_{\rm pl}^4}{[ 2f (r, \theta) ]^2} \left[ {\cal B} (\theta) \mu^2 r^2 + {\cal C} (\theta) \mu r^3 + {\cal D} (\theta) r^4 \right] ,
\label{eqn:VErtheta}
\eeq
in accord with Eq.~(\ref{eqn:VEconformal}). 

\begin{figure*}[t!]
    \includegraphics[width=0.43\textwidth]{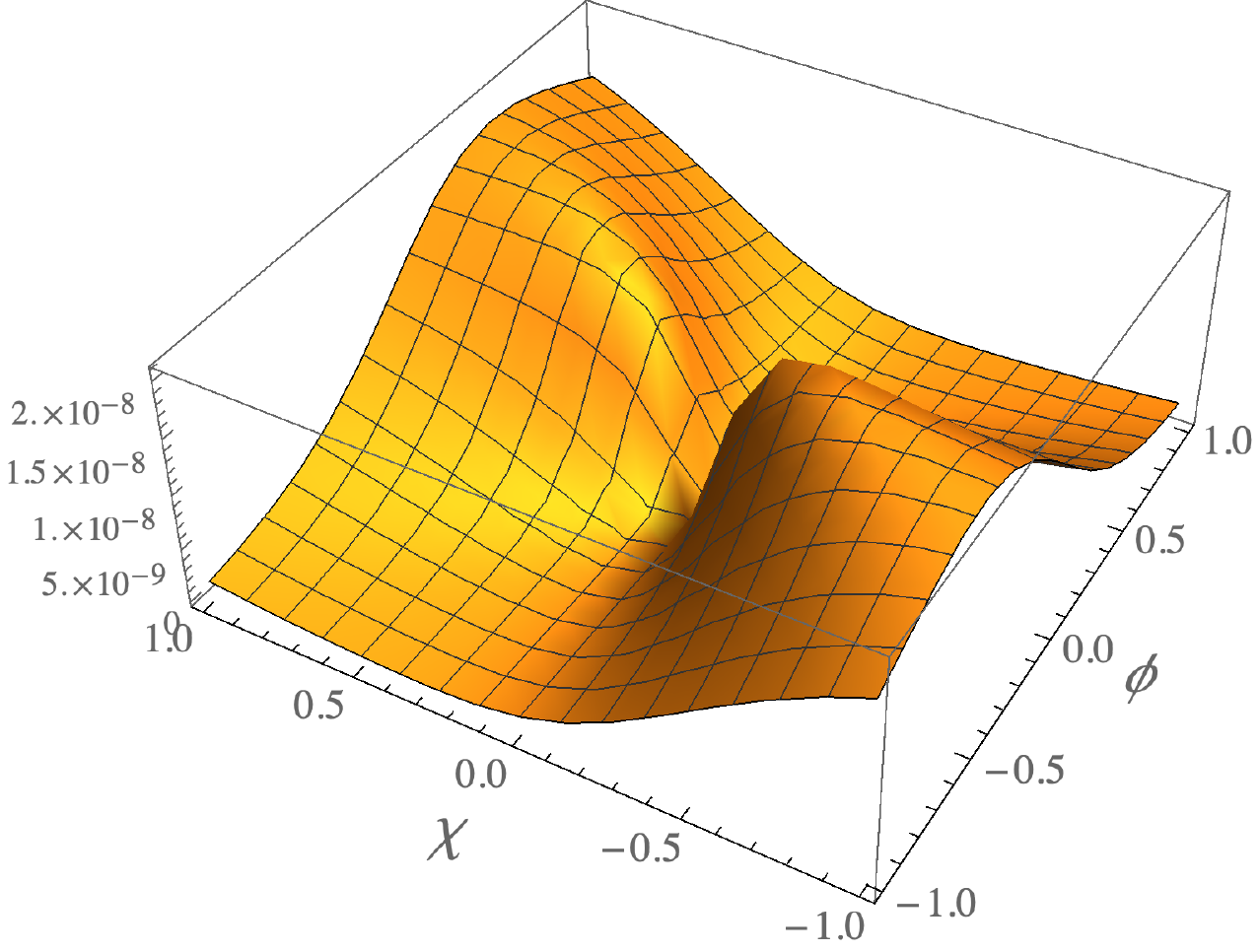} $\quad$ \includegraphics[width=0.43\textwidth]{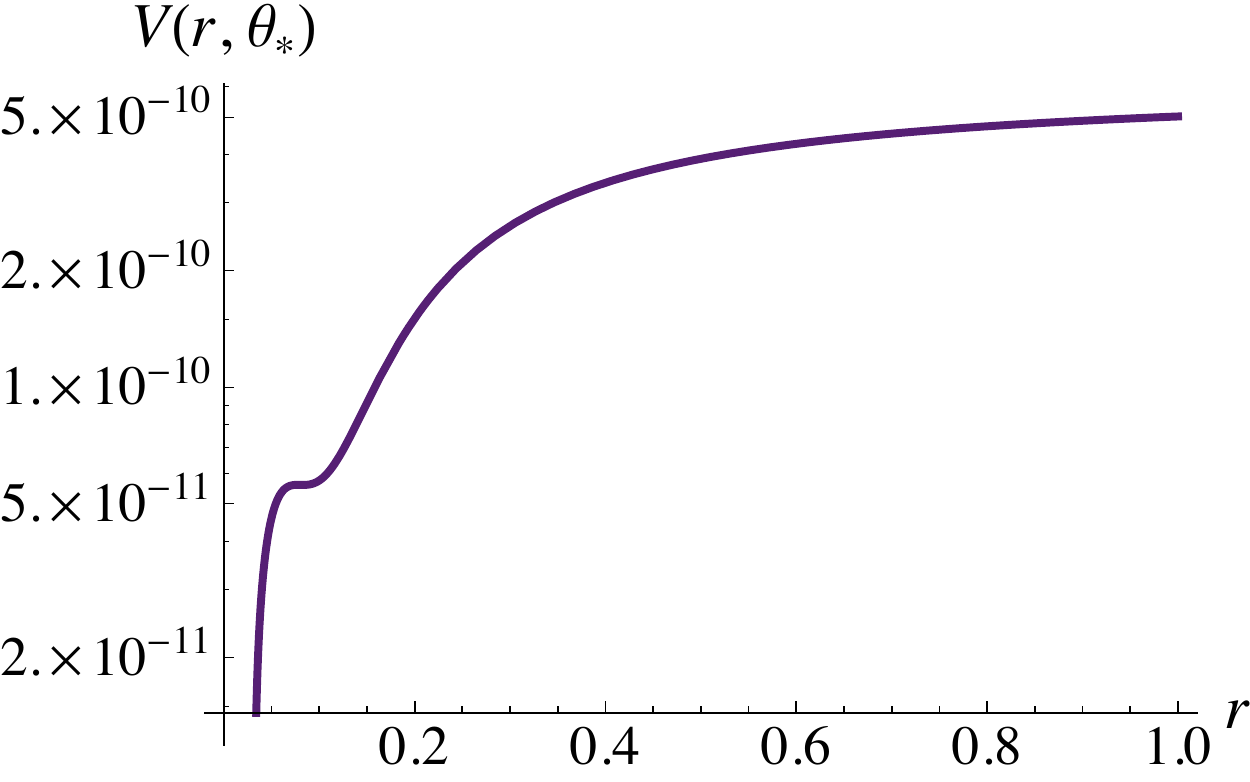}
    \caption{ ({\it Left}) The potential $V (\phi, \chi)$ in the Einstein frame, with fields shown in units of $M_{\rm pl}$. The parameters are $\mu = M_{\rm pl}$, $b = -1.8 \times 10^{-4}$, $c_1 = 2.5 \times 10^{-4}$, $c_2 = 3.5709 \times 10^{-3}$, $c_4 = 3.9 \times 10^{-3}$, and $\xi = 100$. ({\it Right}) The Einstein-frame potential $V (r, \theta_*)$ evaluated along the direction of the fields' evolution, $\theta_* (r)$.
    }
    \label{fig:VE}
\end{figure*}

\begin{figure}[h!]
    \centering
    \includegraphics[width=0.45\textwidth]{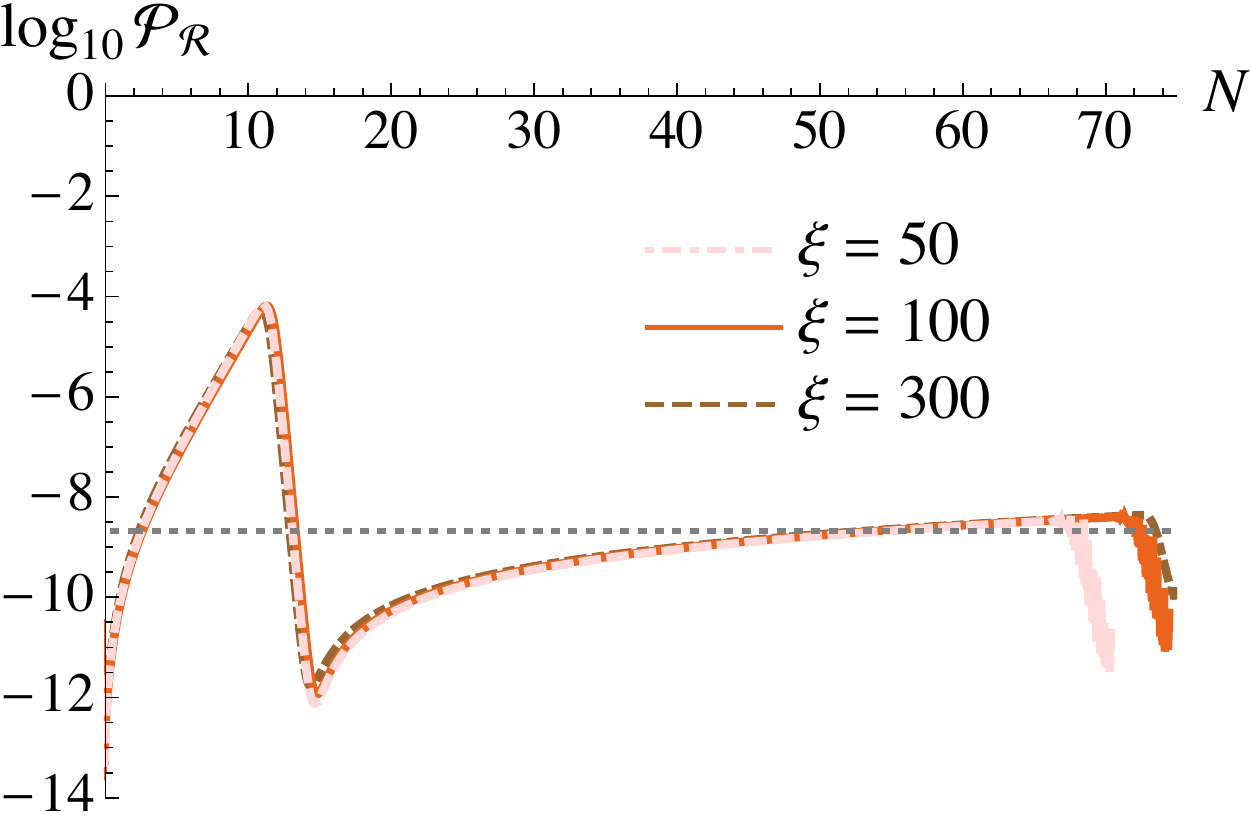}
    \caption{The power spectrum ${\cal P}_{\cal R} (k)$ for three values of the nonminimal coupling constant $\xi$, when we exploit the scaling relationships of Eq.~\eqref{eqn:scaling}. 
    For each curve we set $\mu = M_{\rm pl}$, $\hat{c}_1 = 2.5 \times 10^{-4}$, and $\hat{c}_4 = 3.9 \times 10^{-3}$. 
    For $\xi = 100$ we set $y = 1$, $\hat{b} = - 1.8 \times 10^{-4}$, and $\hat{c}_2 = 3.5709 \times 10^{-3}$.
    For the remaining curves, appropriate values of $y$ and $\hat{b}$ follow from Eq.~(\ref{eqn:scaling}); we also adjust $\hat{c}_2$ in each case by $\Delta \hat{c}_2 / \hat{c}_2 = {\cal O} (10^{-4})$, because the dynamics only become exactly invariant under the scaling of Eq.~(\ref{eqn:scaling}) in the limit $\xi \rightarrow \infty$.
    }
    \label{fig:PRscaling}
\end{figure}

\subsection{Effective Single-Field Evolution}

Whereas such models ostensibly include multiple interacting scalar fields, they generically exhibit dynamics that are {\it effectively} single-field. Following the conformal transformation, the effective potential in the Einstein frame $V (r, \theta)$ generically develops local maxima (``ridges") and local minima (``valleys"). Across a wide range of initial conditions and parameter values, dynamics in models with such potentials typically display an initial transient followed by effectively single-field evolution, along what has been dubbed a ``single-field attractor" \cite{Kaiser:2012ak,Kaiser:2013sna,Schutz:2013fua,Kaiser:2015usz}.

As in Ref.~\cite{Geller:2022nkr}, we impose additional symmetries among the couplings,
\beq
\xi_\phi = \xi_\chi = \xi, \>\> b_1 = b_2 = b, \>\>  c_2 = c_3 .
\label{eqn:bcxisymmetries}
\eeq
Imposing the symmetries of Eq.~(\ref{eqn:bcxisymmetries}) yields at least two benefits: it reduces the dimensionality of the (still large) parameter space to explore, and it enables us to find analytic solutions for the direction in field space $\theta_* (r)$ along which the fields evolve during inflation \cite{Geller:2022nkr}. Moreover, upon imposing the symmetries of Eq.~(\ref{eqn:bcxisymmetries}), the mass parameter $\mu$ only enters the dynamics in the combination $b \mu$, so we may set $\mu = M_{\rm pl}$ without loss of generality. Then the attractor dynamics during inflation---and hence predictions for observables---depend only on six free parameters: the five dimensionless couplings $\{ \xi, b, c_1, c_2, c_4 \}$ and one initial condition for the fields, $r (t_0)$. The only constraint on $r (t_0)$ is that it be large enough to yield sufficient inflation, $r (t_0) \geq 10 \, M_{\rm pl} / \sqrt{ \xi}$ \cite{Kaiser:2012ak,Kaiser:2013sna,Schutz:2013fua,Geller:2022nkr}. (Similar attractor behavior has been identified for other well-studied multifield models, such as hybrid inflation \cite{Clesse:2009ur}.)

Under the symmetry of Eq.~(\ref{eqn:bcxisymmetries}), we also find that the Einstein-frame potential is invariant if we scale the parameters and fields as follows:
\beq
b = \sqrt{y} \, \hat{b}, \quad 
c_i= y\hat{c}_i, \quad
\xi = y\hat{\xi}, \quad
r = \hat{r} / \sqrt{y},
\label{eqn:scaling}
\eeq
where $y > 0$ is a real-valued constant.
As field-space scalars, the potential $V (r, \theta)$ and the angle $\theta_* (r)$ are both invariant under the rescalings of Eq.~(\ref{eqn:scaling}). The metric ${\cal G}_{IJ} (r, \theta)$, on the other hand, is a field-space tensor rather than a scalar, whose components {\it do} transform under these rescalings. The field-space curvature in these models falls as $1/\xi$~\cite{Kaiser:2012ak,McDonough:2020gmn}, and therefore the full inflationary dynamics---which depend on both $V (r, \theta)$ and ${\cal G}_{IJ} (r, \theta)$---become invariant under the rescalings of Eq.~(\ref{eqn:scaling}) in the limit $\xi \rightarrow \infty$.

An example of a potential that yields appropriate inflationary dynamics and can also produce PBHs is given in Fig.~\ref{fig:VE}; the parameters for this example are $\mu = M_{\rm pl}$, $b = -1.8 \times 10^{-4}$, $c_1 = 2.5 \times 10^{-4}$, $c_2 = 3.5709 \times 10^{-3}$, $c_4 = 3.9 \times 10^{-3}$, and $\xi = 100$. The left panel shows the Einstein-frame potential $V (\phi, \chi)$, and the right panel shows the potential $V (r, \theta_*)$ projected along the direction $\theta_* (r)$, corresponding to the single-field attractor evolution.
The effective potential along the attractor direction has a local minimum and nearby local maximum at small field values, set by the condition $\vert {\cal C} (\theta_*) \vert \mu r \sim {\cal B} \mu^2 + {\cal D} (\theta_*) r^2$ \cite{Geller:2022nkr}. Much as in single-field models \cite{Garcia-Bellido:2017mdw,Ezquiaga:2017fvi,Germani:2017bcs,Kannike:2017bxn}, such a feature in the potential can induce a brief phase of ultra-slow-roll evolution, generating a sharp spike in the power spectrum ${\cal P}_{\cal R} (k)$ at specific wavenumbers $k$ \cite{Geller:2022nkr}.

The power spectrum for the set of parameters used in Fig.~\ref{fig:VE} is shown in Fig.~\ref{fig:PRscaling} as the dark orange line.
To demonstrate the scaling relation of Eq.~\eqref{eqn:scaling}, we multiply the fiducial parameter set by appropriate factors of $y$ such that we obtain the scaled parameter sets corresponding to $\xi=50$ and $300$. We further adjust $\hat{c}_2$ by $\Delta \hat{c}_2 / \hat{c}_2 \sim 1/\xi^2 \sim {\cal O} (10^{-4})$ in each case, to accommodate the non-invariance of ${\cal G}_{IJ} (r, \theta)$ under the scaling relations of Eq.~(\ref{eqn:scaling}) for finite $\xi$. The power spectra for these scaled parameters are also shown on Fig.~\ref{fig:PRscaling}; the curves are all nearly identical. Hence, given a set of parameters with some value of $\xi$, we can use the scaling relation to find a corresponding set of parameters at a different value of $\xi$ that will yield the same predictions for observables. We will take advantage of this property to reduce the dimensionality of parameter space explored by the MCMC described in \S~\ref{sec:results}.

\section{Observables}
\label{sec:observables}

Upon imposing the symmetries between couplings given by Eq.~(\ref{eqn:bcxisymmetries}), the two-field models under consideration are specified by five free dimensionless parameters, $\{ \xi, b, c_1, c_2, c_4 \}$, and the fields' initial value $r (t_0)$. Our aim is to determine how generically such models will satisfy CMB constraints, produce PBHs that could account for the DM abundance, and produce detectable GW signals.

We do so by determining the regions of parameter space that yield predictions that are consistent with both the empirical constraints and meet the criteria for producing PBHs. The latter---which yield a population of PBHs within the mass range of interest---are more restrictive, since by slightly relaxing these PBH constraints, the model generally remains in compliance with CMB observational constraints. In this section, we identify specific observables of interest and consider how model predictions for these observables vary with parameters.

\subsection{Cosmic Microwave Background }
\label{CMBsubsection}

The dimensionless power spectrum for the gauge-invariant curvature perturbations, ${\cal P}_{\cal R} (k)$, defined in Eq.~(\ref{PRdefgeneral}), is central to the consideration of CMB constraints. In particular, predictions from this model must be consistent with the latest high-precision measurements of several quantities related to ${\cal P}_{\cal R} (k)$ in the vicinity of the CMB pivot scale $k_* = 0.05 \, {\rm Mpc}^{-1}$ \cite{Planck:2018jri,Planck:2018vyg,Planck:2019kim,BICEP:2021xfz}, including the amplitude (or COBE normalization)
\beq
A_s \equiv {\cal P}_{\cal R} (k_*)  
\label{AsPlanck}
\eeq
and the spectral index
\begin{align}
    n_s (k_*) &\equiv 1 + \left( \frac{d\, {\rm ln} {\cal P}_{\cal R} (k) }{ d \,{\rm ln} k} \right) \Big\vert_{k_*} \n 
    &\simeq 1 - 6 \epsilon (t_*) + 2 \eta (t_*) ,
    \label{nsdef}
    \end{align}
where the second line holds to first order in slow-roll parameters, $\epsilon$ and $\eta$ are defined in Eq.~(\ref{eqn:slowroll}), and $t_*$ is the time when $k_*$ crosses outside of the Hubble radius. We also consider the running of the spectral index,
\begin{equation}
\alpha (k_*) \equiv \left( \frac{ d n_s (k) }{ d \, {\rm ln} k} \right) \Big\vert_{k_*} \simeq \left( \frac{ \dot{n}_s (k)}{H} \right) \Big \vert_{k_*}.
\label{alphadef}
\end{equation}
Observables related to the CMB may be calculated using the expression in Eq.~(\ref{eqn:PRHepsilon}) in our model, across all the regions of parameter space under study here.
    
As noted in the previous section, within these models the fields evolve along single-field attractors during inflation, with exponentially suppressed turning within field space. In the limit $\vert \omega^I \vert \ll H$, the tensor-to-scalar ratio for our multifield models reverts to its usual single-field form \cite{Bassett:2005xm,Gong:2016qmq,Kaiser:2012ak,Geller:2022nkr}
\begin{equation}
    r (k_*) = 16 \epsilon (t_*) .
    \label{rTtoS}
\end{equation}
Given that the isocurvature perturbations remain heavy and the turn rate remains suppressed in these models, we also find that typical multifield features, such as primordial isocurvature perturbations $\beta_{\rm iso} (k_*)$ and primordial non-Gaussianity (parameterized by various dimensionless coefficients $f_{\rm NL}$, corresponding to different shape functions for the bispectrum) remain exponentially suppressed \cite{Kaiser:2013sna,Kaiser:2015usz,Geller:2022nkr}, easily consistent with the latest observations \cite{Planck:2018jri,Planck:2019kim}.

We compare predictions from our model with the {\it Planck} 2018 results (when the spectral index is allowed to run with wavenumber)~\cite{Planck:2018jri} and the {\it Planck}-BICEP/Keck 2021 constraint on the tensor-to-scalar ratio \cite{BICEP:2021xfz}:
\beq
\begin{split}
    \ln (10^{10} A_s ) &= 3.044 \pm 0.014 , \\
    n_s (k_*) &= 0.9625 \pm 0.0048 , \\
    \alpha_s (k_*) &= 0.002 \pm 0.010 , \\
    r (k_*) &< 0.036 ,
    \label{eqn:PlanckCMBconstraints}
\end{split}
\eeq
where the reported error bars correspond to $68\%$ confidence intervals.

\subsection{Primordial Black Holes }
\label{sec:PBHconstraints}

The primordial power spectrum ${\cal P}_{\cal R} (k)$ must exceed some threshold on appropriate scales $k_{\rm pbh}$ in order for the curvature perturbations to seed primordial overdensities that will ultimately undergo gravitational collapse when these perturbations re-enter the Hubble radius after the end of inflation. For the models under consideration, this threshold is achieved for modes with $k=k_{\text{pbh}}$ that cross outside the Hubble radius during the transient period of ultra-slow roll.
Typical estimates suggest a threshold for PBH formation of ${\cal P}_{\cal R} (k_{\rm pbh}) \geq 10^{-3}$, about six orders of magnitude greater than the amplitude around the CMB pivot scale $k_*$, ${\cal P}_{\cal R}  (k_*) = A_s = 2.1 \times 10^{-9}$ \cite{Kuhnel:2015vtw,Young:2019yug,Kehagias:2019eil,Escriva:2019phb,DeLuca:2020ioi,Musco:2020jjb,Escriva:2021aeh}. 

Various effects beyond linear order in perturbation theory, such as stochastic dynamics and quantum diffusion, typically yield a non-Gaussian probability distribution function for curvature perturbations of various amplitudes, increasing the likelihood of large-amplitude perturbations compared to the Gaussian approximation. Such effects, in turn, can reduce the required threshold on ${\cal P}_{\cal R} (k_{\rm pbh})$ by one to two orders of magnitude \cite{Ezquiaga:2019ftu,Figueroa:2020jkf,Tada:2021zzj,Biagetti:2021eep,Animali:2022otk,Gow:2022jfb,Ferrante:2022mui}. Nevertheless, in this work we use the threshold ${\cal P}_{\cal R} (k_{\rm pbh}) \geq 10^{-3}$; this is conservative in that relaxing this threshold would only lead to a larger region of parameter space that would be consistent with observations.
For computational tractability, we impose this conservative threshold via ${\cal P}_{\cal R}^{\rm SR} (k_{\rm pbh}) \geq 10^{-3}$, given that ${\cal P}_{\cal R} (k) \geq {\cal P}_{\cal R}^{\rm SR} (k)$ when ultra-slow-roll effects are taken into account, as discussed further in Appendix \ref{app:USR}.

In addition to the peak height of the power spectrum, the PBHs that form after inflation are also sensitive to the time, during inflation, when the large-amplitude perturbations were first amplified and crossed outside the Hubble radius. We denote this time as 
\begin{align}
\Delta N \equiv N_{\rm pbh} - N_{\rm end},
\end{align}
where $\Delta N$ is the number of $e$-folds before the end of inflation when $k_{\rm pbh}$ crossed outside the Hubble radius.

To determine an appropriate range for $\Delta N$, we note that after the end of inflation, when perturbations of comoving wavenumber $k_{\rm pbh}$ cross back inside the Hubble radius and induce gravitational collapse at time $t_c$, the peak of the mass distribution $M_{\rm pbh} (t_c)$ scales with the Hubble mass $M_H (t_c)$ as $M_{\rm pbh} (t_c) = \gamma \, M_H (t_c)$, with $\gamma \simeq 0.2$. (Here $M_H (t_c)\equiv 4\pi \rho (t_c) / (3 H^3 (t_c))$, where $\rho (t_c)$ is the energy density contained with a Hubble sphere of radius $H^{-1} (t_c)$.) One may then relate $k_{\rm pbh}$ to $M_{\rm pbh}$ \cite{Ozsoy:2023ryl}
\beq
\frac{ k_{\rm pbh} }{3.2 \times 10^5 \, {\rm Mpc}^{-1}} =  \left( \frac{ 30 \, M_\odot} {M_{\rm pbh}} \right)^{1/2} \left( \frac{ \gamma}{0.2} \right)^{1/2} \left( \frac{ g_* (T_c) }{106.75} \right)^{-1/12} ,
\label{kpbhMsun}
\eeq
where $g_* (T_c)$ is the number of effectively massless degrees of freedom at the time of PBH formation. PBHs with $M_{\rm pbh}$ in the range of interest for DM, $10^{17} \, {\rm g} \leq M_{\rm pbh} \leq 10^{22} \, {\rm g}$ \cite{Carr:2020gox,Carr:2020xqk,Green:2020jor,Villanueva-Domingo:2021spv,Escriva:2021aeh,Escriva:2022duf,Ozsoy:2023ryl}, will therefore form from the collapse of perturbations with comoving wavenumber $10^{11} \, {\rm Mpc}^{-1} \leq k_{\rm pbh} \leq 10^{14} \, {\rm Mpc}^{-1}$. 

Next we evaluate the time between when the CMB pivot scale $k_*= 0.05 \, {\rm Mpc}^{-1}$ and modes of wavenumber $k_{\rm pbh}$ each first crossed outside the Hubble radius:
\beq
\frac{ k_{\rm pbh} }{k_*} = \frac{ a (t_{\rm pbh} )}{a (t_*)} \frac{ H (t_{\rm pbh} )}{ H (t_*)} = \left( \frac{ H (t_{\rm pbh} )}{H (t_*)}\right) e^{N_* - \Delta N } ,
\label{kcompare}
\eeq
where $N_*$ corresponds to the number of $e$-folds before the end of inflation at which the pivot scale crossed the Hubble radius.
In our models, the Hubble rate falls between $t_*$ and $t_{\rm pbh}$, with $H (t_{\rm pbh}) / H(t_*) \leq 1/2$ within the region of parameter space of interest. Taking into account the residual uncertainty on the duration of the reheating period \cite{Amin:2014eta} we set $N_* = 55 \pm 5$, which yields 
\beq
\Delta N \geq 14
\label{DeltaN14}
\eeq
$e$-folds before the end of inflation. 

As discussed in Appendix \ref{app:USR}, setting $\Delta N \geq 14$ is a conservative threshold, in that additional effects can only increase $\Delta N$. For example, growth during the ultra-slow-roll phase modifies the peak wavenumber $k_{\rm pbh}$, yielding $k_{\rm pbh}^{\rm USR} \leq k_{\rm pbh}^{\rm SR}$ (and hence $\Delta N^{\rm USR} \geq \Delta N^{\rm SR})$, where the ``SR" quantities are evaluated in terms of the slow-roll expression in Eq.~(\ref{eqn:PRHepsilon}). Given these effects, in addition to the uncertainties from non-Gaussian effects and the reheating phase noted above, it is premature to plot a distribution of the resulting PBH masses or $k_\textrm{pbh}$, since an uncertainty that increases $\Delta N$ by a few $e$-folds reduces $k_\textrm{pbh}$ as per Eq.~\eqref{kcompare}, which in turn increases $M_{\rm pbh}$ as per Eq.~\eqref{kpbhMsun}. Instead, by imposing Eq.~(\ref{DeltaN14}), we ensure that the region of our resulting parameter space that passes the threshold will produce PBHs large enough to avoid evaporation bounds.

\subsection{Parameter Dependence \& Degeneracies}
\label{sec:degen1}
%
\begin{figure*}[t!]
    \includegraphics[width=0.49\textwidth]{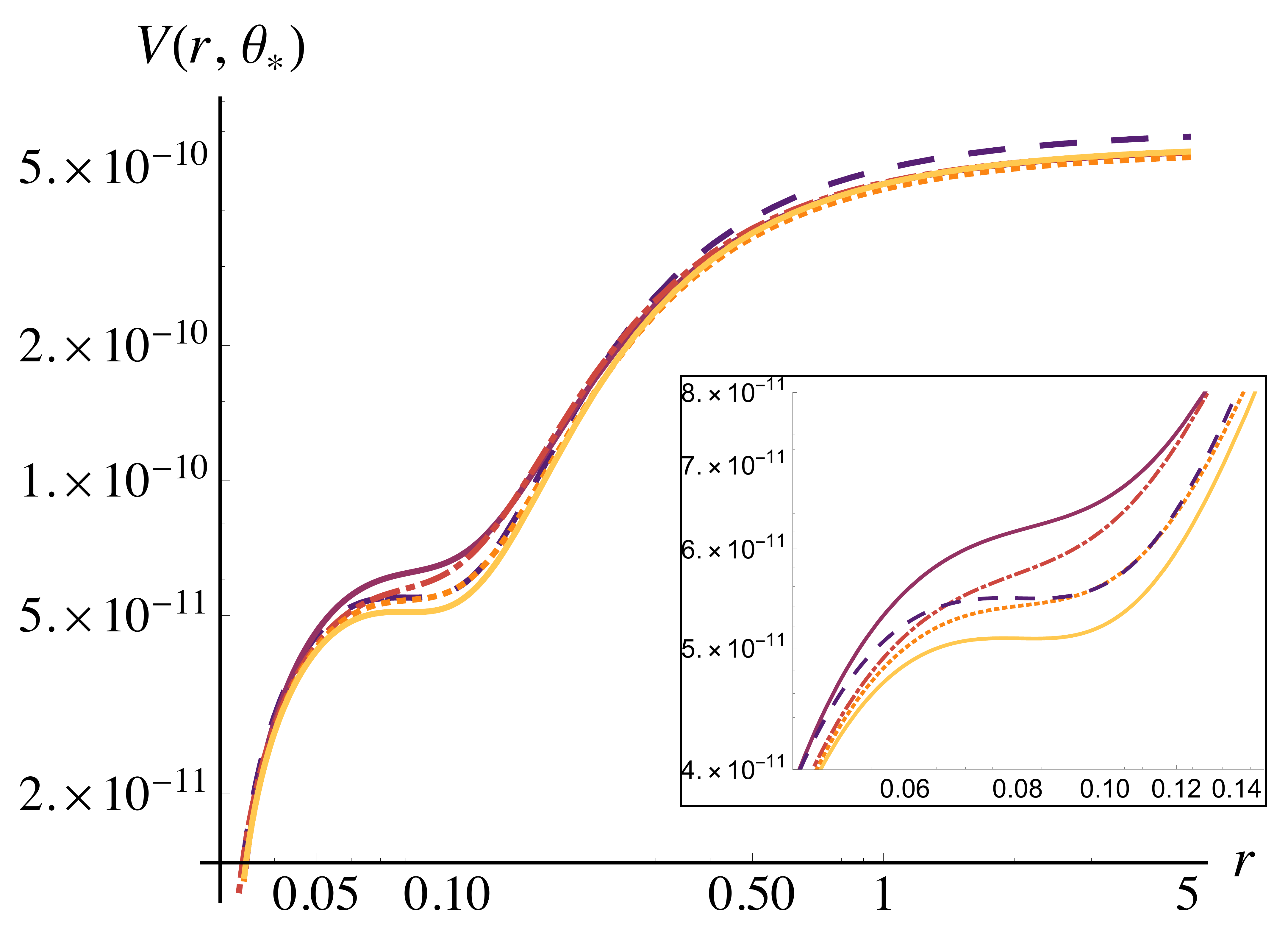}\quad
    \includegraphics[width=0.48\textwidth]{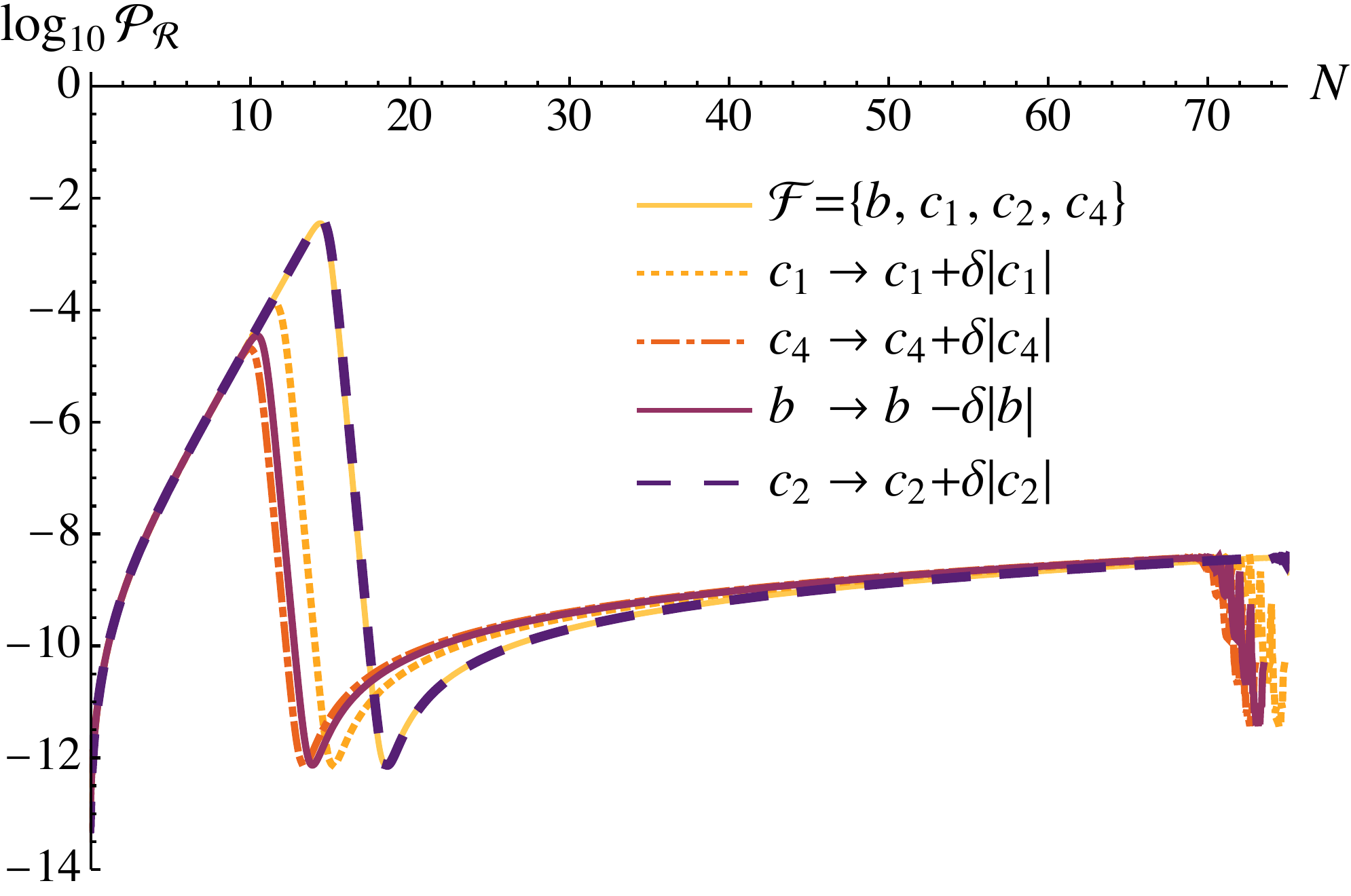}
    \caption{
    The potential (\emph{left panel}) and power spectrum (\emph{right panel}) are plotted for parameter set $\mathcal{F}=\{b, c_1, c_2, c_4\}$ in yellow. We then vary one parameter at a time, cumulatively, to obtain each of the other curves: first we increase $c_1$ (orange dotted curve), then increase $c_4$  (red dot-dashed curve), then decrease $b$ (magenta curve), and finally increase $c_2$ to obtain a degenerate parameter set (dashed purple curve), $\mathcal{F}'=\{b-\delta|b|, c_1+\delta|c_1|, c_2+\delta|c_2|, c_4+\delta|c_4|\}$. 
    The power spectrum is much more sensitive to the step size along the degeneracy direction than the potential; hence, the parameter variations in the left panel are of order $\delta\simeq 10^{-2}$, whereas in the right panel they are of order $\delta\simeq 10^{-6}$.
    }
    \label{fig:V_PR_degen}
\end{figure*}

Once the nonminimal coupling $\xi$ is fixed, the parameter space is described by the four remaining free parameters, $b, c_1, c_2, c_4$. Varying these parameters changes the behavior of the potential and thus also changes predictions for $n_s$, $\mathcal{P}_{\mathcal{R}}$, and other observables in characteristic ways.

Measurements of CMB observables are sensitive to physics at the pivot scale, thus they will be affected by changes to the potential at \textit{large} field values, that is, around $r(t_*)$. Meanwhile, the PBH constraints are largely sensitive to changes in the potential at \textit{small} field values, around $r(t_{\text{pbh}})$, corresponding to the period of ultra-slow roll during which the modes with $k\sim k_{\text{pbh}}$ first exit the Hubble radius. The tension is thus between tuning the small-field features to get a sufficiently large spike in $\mathcal{P}_{\mathcal{R}}$ to seed PBH formation without compromising the large-field dynamics. 
(See also Refs.~\cite{Geller:2022nkr,Byrnes:2018txb,Carrilho:2019oqg,Ando:2020fjm}.)

In general, the longer the period of ultra-slow roll, that is, the larger the relative depth of the local minimum at small field values as compared to the local maximum and the large-field plateau, the larger the spike in the primordial power spectrum $\mathcal{P}_{\mathcal{R}}$ will be. There is a limit to this trend, however. If the relative depth of the local minimum is too large, the characteristic time for the fields to quantum tunnel out of the local minimum becomes comparable to the time for classical transit through the ultra-slow-roll region. In this instance, we cannot ignore the effects of quantum diffusion on the dynamics, hence, there is a range of parameters for which the fields undergo ultra-slow roll evolution long enough for PBHs to form post-inflation, but not so long that quantum effects become dominant. 
We can determine whether a potential will have a small-field feature that falls within this range by considering the magnitude of kinetic energy for the fields as they enter the ultra-slow-roll region. If the kinetic energy is too high, the fields will roll past the region of the potential for which $V_{,\sigma}\simeq 0$ too quickly for ultra-slow roll to yield a sufficient spike in $\mathcal{P}_{\mathcal{R}}$, whereas if the kinetic energy is too low, the fields will become classically ``stuck'' in the local minimum. 
This is discussed in more detail in our previous work \cite{Geller:2022nkr}. 

Varying each of the parameters individually affects both the shape of the minimum/maximum feature at small field values as well as the slope of the potential between the large-field plateau and the local minimum, which will change the kinetic energy of the fields as they approach the region of the potential for which $V_{,\sigma}\simeq 0$, as follows: 

\begin{itemize}
    \item Increasing $|b|$ while keeping $b<0$ will both increase the relative depth of the local minimum while \textit{decreasing} the slope of the potential as the fields approach the region with $V_{,\sigma}\simeq 0$. The overall effect is to \emph{decrease} the kinetic energy of the fields as they enter the ultra-slow roll regime.
    
    \item Increasing $c_2=|c_2|$ also increases the relative depth of the local minimum, but \emph{increases} the slope of the potential as the fields approach $V_{,\sigma}\simeq 0$. The latter effect dominates, so the net result is to \textit{increase} the kinetic energy of the fields. 
    
    \item Increasing both $c_1=|c_1|$ and $c_4=|c_4|$ will \textit{increase} the kinetic energy of the fields as they approach the ultra-slow roll region. 
\end{itemize}

Thus the effect of parameter variations is such that increasing the magnitude of $b$ has the opposite effect to increasing the magnitudes of the $c_i$. We find that the interplay of parameters is such that a certain sequence of small parameter variations will lead to a degenerate potential and power spectrum. An example of this is shown in Fig.~\ref{fig:V_PR_degen}. We begin at the fiducial parameter set $\vec{\mathcal{F}}$ and take a small step $\delta$ in the degeneracy direction given by the unit vector $\hat{n}$ to a degenerate parameter set $\vec{\mathcal{F}}^{'}=\vec{\mathcal{F}}+\hat{n}\delta$, where $\hat{n}\delta=(-|b|,|c_1|,| c_2|, |c_4|)\delta$. The power spectrum is much more sensitive to the step size along the degeneracy direction than is the potential, so the right panel, for ${\cal P}_{\cal R} (k)$, uses $\delta\simeq 10^{-6}$, whereas the left panel, for $V (r, \theta_*)$, uses a step size of $\delta\simeq 10^{-2}$ so that the effects of varying parameters can be more readily seen.

\subsection{Gravitational Waves}
\label{sec:GWs}

In this work we also consider a complementary observable in inflationary PBH models, namely GWs sourced by the amplified curvature perturbation.

At second order in cosmological perturbation theory, scalar modes can source a SGWB. The induced GWs are usually described by the energy density $\rho_{\rm GW}$ per logarithmic frequency interval, normalized by the critical density (see, e.g., Ref.~\cite{Domenech:2021ztg})
\begin{equation}
    \Omega_{\textrm{GW,0}} h^2 = \frac{h^2}{3M_{\rm Pl}^2 H_0^2} \frac{d\rho_{\rm GW}}{d \textrm{ln} k} .
\end{equation}
Assuming that the GWs were induced by modes that crossed back inside the Hubble radius at a temperature $T_c$ during the radiation-dominated epoch, the fractional energy density today can be written as
\begin{equation}
    \Omega_{\textrm{GW,0}} h^2 = \Omega_{r0}h^2 \left( \frac{g_* (T_c)}{g_{*,0}} \right) \left( \frac{g_{*s} (T_c)}{g_{*s,0}} \right)^{-4/3} \Omega_{\textrm{GW,c}} , 
\end{equation}
where $H_0$ is the present value of the Hubble constant, $h = H_0 / (100 \,\mathrm{km}\, \mathrm{s}^{-1}\, \mathrm{Mpc}^{-1})$, and $\Omega_{\textrm{GW,c}}$ is the GW spectral density at the time the waves were induced. The quantities
$g_* (T)$ and $g_{*s} (T)$ are the effective number of degrees of freedom for the radiation energy density and entropy; today, their values are equal to $g_{*,0} = 3.36$ and $g_{*,0} = 3.91$.
Using $\Omega_{r0}h^2 = 4.18 \times 10^{-5}$ \cite{Planck:2018vyg} and $g_* (T_c) \approx g_{*s} (T_c) \approx 106.75$, this becomes 
\begin{align}
\label{eq:dimensionlessGWspectraldensity}
    \Omega_{\textrm{GW,0}} h^2 \approx 1.62 \times 10^{-5} \, \Omega_{\textrm{GW,c}}.
\end{align}

The dimensionless spectral density when the modes are contained within the Hubble radius during the radiation-dominated epoch is given by
\begin{align}
    \label{eq:GWspectrum}
    \Omega_{\textrm{GW,c}}(k,\tau)&=\frac{1}{24}\left(\frac{k}{aH}\right)^2 \overline{P_h(k,\tau)},
\end{align}
where the conformal time is defined as $\tau = (aH)^{-1}$ at horizon reentry in the radiation-dominated era, and the two respective polarization modes of GWs have been summed over. $P_h$ is the power spectrum of the induced tensor-mode perturbation sourced by linear scalar-mode perturbations at second order given by Eq.~ \eqref{def_P_h}, which can be solved via the Green's function method \cite{Ananda:2006af,Baumann:2007zm} as
\begin{align}\label{sol_h_Green}
    h_\lambda(\vec{k},\tau) = 4 \int^{\tau} d\tau_1 G_{\vec{k}}(\tau;\tau_1) \frac{a(\tau_1)}{a(\tau)} S_\lambda (\vec{k},\tau_1),
\end{align}
where $\lambda = +, \times$ are the two polarizations, $G_{\vec{k}}(\tau;\tau_1) =\frac{1}{k}\sin(k(\tau -\tau_1))$ is the Green's function in radiation domination, and $S_\lambda$ is the source term; detailed information is provided in Appendix~\ref{app:tensor_spectrum}.
The overline in Eq.~\eqref{eq:GWspectrum} denotes an average over a few wavelengths for time oscillations led by the Green's function.
The GW spectrum induced by curvature perturbations is given by~\cite{Kohri:2018awv,Domenech:2021ztg}
\begin{align}
    P_h (\tau, k) = &2 \int_0^\infty \,dt \int_{-1}^1 \,ds \left[ \frac{t (2+t) (s^2-1)}{(1-s+t) (1+s+t)} \right]^2 \n
    &\times I^2 (v, u, x) \mathcal{P_R}(kv) \mathcal{P_R}(ku),
    \label{eqn:Ph}
\end{align}
where $u = (t+s+1)/2$, $v = (t-s+1)/2$, $x = k\eta$, and the appropriate kernel $I (v, u, x)$ is given in Appendix~\ref{app:tensor_spectrum}.

\section{Planck Constraints on PBH-seeding Multifield Inflation}
\label{sec:constraints}

\subsection{Data and Likelihood}
\label{sec:data}

To constrain the model presented above, we use data from the {\it Planck} 2018 CMB temperature and polarization anisotropies and lensing spectra, and enforce a minimal requirement that the model can produce PBHs that could comprise all DM.
We incorporate the CMB data using Gaussian priors corresponding to the {\it Planck} 2018 constraints on the $\Lambda$CDM cosmological model. Specifically, we use measurements of the spectral index $n_s (k_*)$, the amplitude $\ln\left[10^{10} A_s(k_*)\right]$, and the running of the spectral index  $\alpha (k_*)$, corresponding to the marginalized parameter constraints in the context of the $\Lambda$CDM model. We also enforce a one-sided Gaussian constraint on the tensor-to-scalar ratio $r (k_*)$ corresponding to the combined {\it Planck}-BICEP/Keck observations. The best fit values for these quantities and their error bars are given in Eq.~\eqref{eqn:PlanckCMBconstraints}.

In order to translate these constraints to an inflation model, one must assume a reheating scenario; the CMB measurements are reported at the pivot scale $k_*$, and calculating the time during inflation when this mode crossed the Hubble radius requires knowing the time spent during the reheating phase~\cite{Dodelson:2003vq,Liddle:2003as,Amin:2014eta,Martin:2021frd}.
If we assume that reheating is efficient and lasts for much less than one e-fold, then the CMB pivot scale for this model with typical parameters corresponds to $N_* \simeq 58$~\cite{Geller:2022nkr}; however, the longer the reheating phase lasts, the smaller $N_*$ is. 
In inflationary models similar to the one that we consider, post-inflation reheating is typically efficient and lasts for $N_\text{reh} \sim \mathcal{O} (1)$ $e$-folds \cite{Bezrukov:2008ut,Garcia-Bellido:2008ycs,Child:2013ria,DeCross:2015uza,DeCross:2016fdz,DeCross:2016cbs,Figueroa:2016dsc,Repond:2016sol,Ema:2016dny,Sfakianakis:2018lzf,Nguyen:2019kbm,vandeVis:2020qcp,Iarygina:2020dwe,Ema:2021xhq,Figueroa:2021iwm,Dux:2022kuk}.

In our analysis, we allow $N_*$ to take on values within the range typically considered \cite{Planck:2018jri}, $N_* = 55 \pm 5$, and fix $N_*$ to optimize our reheat history. In other words, we choose $N_*$ to be the value between 50 and 60 such that the CMB observables at that scale most closely match the measurements listed in Eq.~(\ref{eqn:PlanckCMBconstraints}). An alternative approach would be to marginalize over the possible reheat histories; however, since this would make our MCMC computationally expensive, we choose to fix $N_*$ using this simpler procedure, wherein $N_*$ is treated as a derived parameter parameterizing the optimal reheating scenario. We leave a dedicated study of reheating dependence, e.g., in analogy to Ref.~\cite{PhysRevD.93.103532}, to future work.

To summarize, we take the model likelihood to be the following:
\begin{enumerate}
    \item A Gaussian over the {\it Planck} and BICEP/Keck observables, $\{ \ln(10^{10} A_s), n_s (k_*), \alpha (k_*), r(k_*) \}$, corresponding to the {\it Planck} constraints on each of these quantities.
    
    \item A uniform likelihood for the peak of the power spectrum in the restricted range ${\cal P}_{\cal R} (k_{\rm pbh}) \geq 10^{-3}$, namely the threshold to form PBHs, and zero likelihood for the peak falling below this.
    
    \item A uniform likelihood for the {\it position} of the peak of the power spectrum, in the restricted range $14 \leq \Delta N \leq 25$, corresponding to the mass window where PBHs can comprise an $\mathcal{O}(1)$ fraction of DM, and zero likelihood outside of this range.
\end{enumerate}

Next, in order to determine the observability of the induced GWs from curvature perturbations, we compare our predicted signals to the sensitivity curves from LISA~\cite{Schmitz:2020syl}, LIGO A+~\cite{Cahillane:2022pqm}, the ET~\cite{Maggiore:2019uih}, CE~\cite{Reitze:2019iox}, and DECIGO~\cite{Yagi:2011wg,Kawamura:2020pcg}.
\begin{figure}
	\includegraphics[scale=0.5]{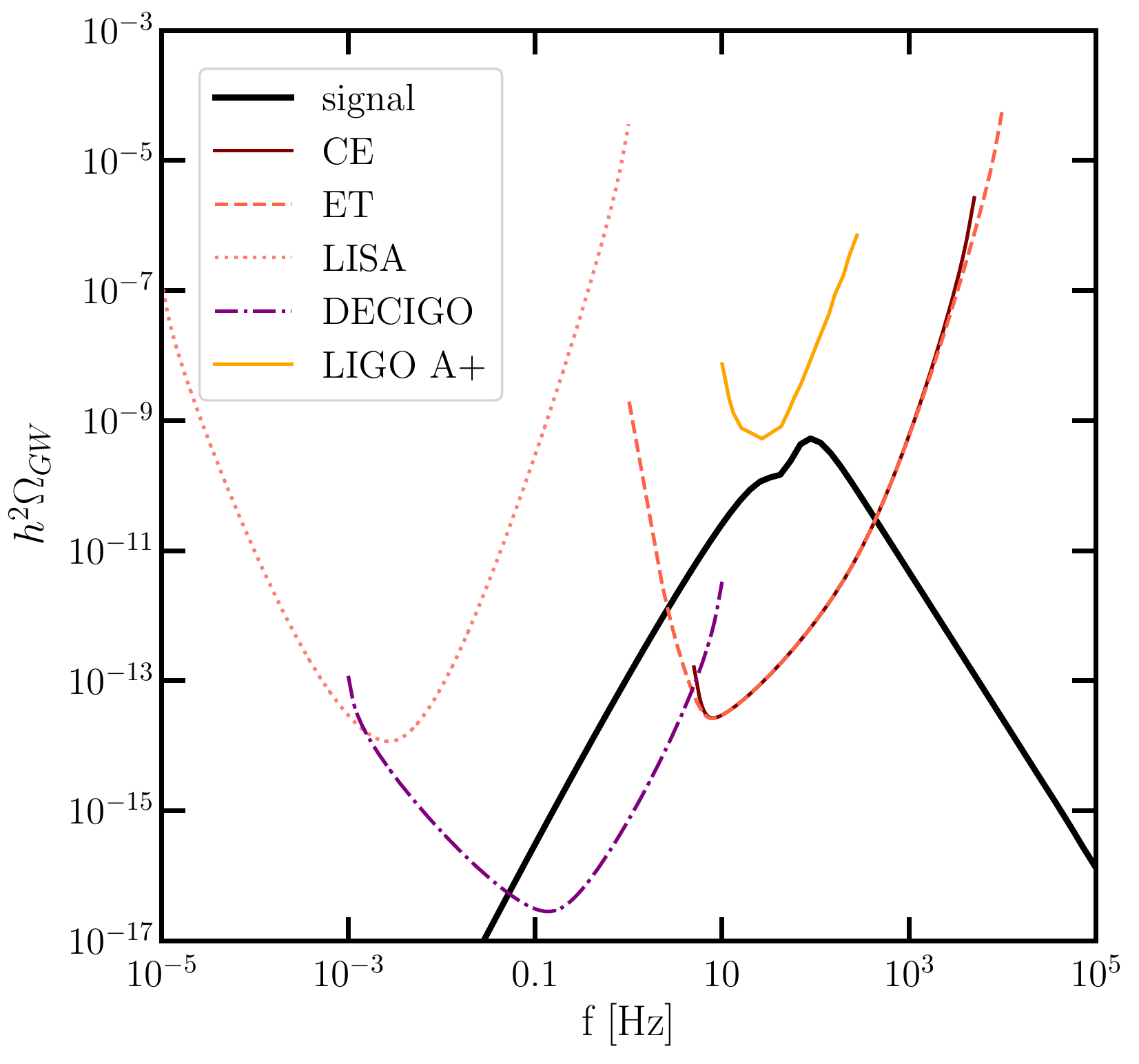}
	\caption{
	The spectral density of gravitational waves from our model with parameters $\xi=100$, $b=-1.8 \times 10^{-4}$, $c_1=2.5 \times 10^{-4}$, $c_2=3.570913 \times 10^{-3}$, and $c_4=3.9 \times 10^{-3}$.
	For comparison, we also plot the power-law integrated sensitivity curves for LIGO A+~\cite{Cahillane:2022pqm}, LISA~\cite{Schmitz:2020syl}, ET~\cite{Maggiore:2019uih}, CE~\cite{Reitze:2019iox}, and DECIGO~\cite{Yagi:2011wg,Kawamura:2020pcg}.}
	\label{fig:SGWB}
\end{figure}
In Fig.~\ref{fig:SGWB}, we show an example of a GW signal from our model with a particular set of parameters against these sensitivity curves.\footnote{Note that for plotting experimental sensitivities, we use power-law integrated sensitivity curves~\cite{Thrane:2013oya}. 
The power-law integrated curves will overlap with a SGWB when the signal-to-noise ratio (SNR) is greater than 1. Hence they are a better tool for visualizing the observability of a signal than the typically reported noise spectra.}
Given the total observation time $t_\text{obs}$ and noise spectrum of an experiment $\Omega_\text{noise} (f)$, we can calculate the signal-to-noise ratio (SNR) of a background of GWs with power spectral density $\Omega_\text{GW,0} (f)$:
\begin{equation}
\label{eq:SNR}
    \rho = \sqrt{2 t_\text{obs} \int_{f_\text{min}}^{f_\text{max}} df \, \left(\frac{\Omega_\text{GW,0} (f)}{\Omega_\text{noise} (f)}\right)^2 }
\end{equation}
The SNR for this signal is maximal for CE at $\rho=1811$ and also quite large for ET at $\rho=667$, whereas $\rho\simeq0$ in LIGO A+ and LISA, and finally in DECIGO the SNR is sizeable at $\rho=243$. 

The frequency limits for integration in the above expression, denoted $[f_\textrm{min},f_\textrm{max}]$, define the bandwidth of the detector. 
Eq.~(\ref{eq:SNR}) therefore represents the total broadband SNR, integrated over both time and frequency. 
It can be computed as the expected SNR of a filtered cross-correlation. 
Here, in general one assumes that the SGWB is sufficiently described by a power-law of the from $\Omega_\textrm{GW}$ = $\Omega_\beta (f /f_\textrm{ref})^\beta$ where $\beta$ is the spectral index and $f_\textrm{ref}$ is the reference frequency, which we set for example to $100$ Hz for ground-based detectors over the sensitivity region of interest. We set the observation times to the duration of data taking by the experiment. We can then use Eq.~\eqref{eq:SNR} to compute the value of GW amplitude required to reach a target SNR. In order to determine the detectability of the SGWB signal, we consider the spectrum from a particular inflationary model to be observable if it gives an SNR of $\rho \geq 1$.

\subsection{Results}
\label{sec:results}

We perform an MCMC analysis \cite{2013PASP..125..306F} of our multifield inflation model fit to cosmological data as described in Sec.~\ref{sec:data}. The posterior sampling is performed using an ensemble sampler \cite{2010CAMCS...5...65G} implemented in the Python package \texttt{emcee} \cite{2013PASP..125..306F}, with 200 walkers. We use the Python package \texttt{Corner} \cite{corner} for plotting results.

Given the scaling relationship in Eq.~\eqref{eqn:scaling}, we choose to fix $\xi =100$ and allow the remaining parameters $b$, $c_1$, $c_2$, and $c_4$ to vary.\footnote{Whereas the dynamics of these models become independent of $\xi$ in the limit $\xi \rightarrow \infty$, we expect that data would not be able to constrain $\xi$ due to the scaling relations of Eq.~(\ref{eqn:scaling}), and the relative constraints on the other parameters would be comparable for any fixed value of $\xi$. Hence we choose to fix $\xi$ to 100.}
We take broad uniform priors on the model parameters given by $b=[-10^{-3},-10^{-4}]$, $c_1 = [10^{-4},10^{-3}]$, $c_2 = [10^{-3},10^{-2}]$, $c_4 = [10^{-3}, 10^{-2}]$.

We assess convergence of our MCMC chains by a combination of the autocorrelation time~\cite{2010CAMCS...5...65G,2013PASP..125..306F} and stability of marginalized parameter constraints.
The \texttt{emcee} documentation recommends running an analysis for 50 autocorrelation times to ensure convergence; however, this would be prohibitively computationally expensive for our case. On the other hand, as also noted in Ref.~\cite{2013PASP..125..306F}, an accurate approximation to marginalized parameter constraints can be realized with significantly fewer samples.  
In total, we include approximately 1,300,000 samples for the final analysis, corresponding to an estimated 11 autocorrelation times. We find that as we vary the number of samples included in the analysis by 10\%, the marginalized parameter constraints (central value and error bars) vary at the sub-percent level.

\begin{table}[htb!]
  Constraints from requiring PBH DM \\
  and satisfying \emph{Planck} 2018 data \\
  \renewcommand{\arraystretch}{1.5}
  \begin{tabular}{|l|c|}
    \hline\hline
    Parameter & Constraint \\ 
    \hline \hline
    
    {\boldmath$b$} & $-1.87\, (-1.73)_{-0.11}^{+0.09} \times 10^{-4}$  \\
    {\boldmath$c_1$} & $2.61 \,(2.34) _{-0.17}^{+0.24} \times 10^{-4}$  \\
    {\boldmath$c_2$} & $3.69\, (3.42) _{-0.16}^{+0.22} \times 10^{-3}$ \\
    {\boldmath$c_4$} & $4.03\, (3.75) _{-0.17}^{+0.24} \times 10^{-3}$ \\

    \hline
    
    $n_s (k_*)$ & $0.952 \, (0.956) _{-0.003}^{+0.002}$   \\
    $\ln (10^{10} A_s)$ & $3.049 \, (3.048) _{-0.001}^{+0.001}$  \\
     $N_*$ & $58.8\, (60.0) _{-2.2}^{+1.2}$  \\
     $\alpha (k_*)$ & $-0.0012\, (-0.0010) _{-0.0002}^{+0.0001}$  \\
     $r (k_*)$ & $0.019 \, (0.016) _{-0.001}^{+0.002}$ \\

    \hline

    $b/c_2$ & $-5.04 (-5.05) _{-0.05}^{+0.03} \times 10^{-2}$  \\
    $c_1/c_2$ & $7.07 (6.84) _{-0.26}^{+0.32} \times 10^{-2}$  \\
    $c_4/c_2$ & $1.091 (1.096) _{-0.008}^{+0.009}$  \\

    \hline
  \end{tabular} 
  \caption{
  The mean (best-fit) $\pm1\sigma$ constraints on parameters and derived observables of our multifield inflation model. Sampled parameters are in boldface.
  To generate these constraints, we fix $\xi=100$, require that the peak of the curvature power spectrum satisfies the requirements to produce PBH DM, and also include \emph{Planck} 2018 measurements on $A_s$, $n_s$, $\alpha$ and $r$.
  The last three rows of the table show the constraints on ratios of model parameters, which are subject to fewer degeneracies.
  }
  \label{tab:best_fit_params}
\end{table}

\begin{figure*}
    \includegraphics[width=0.8\textwidth]{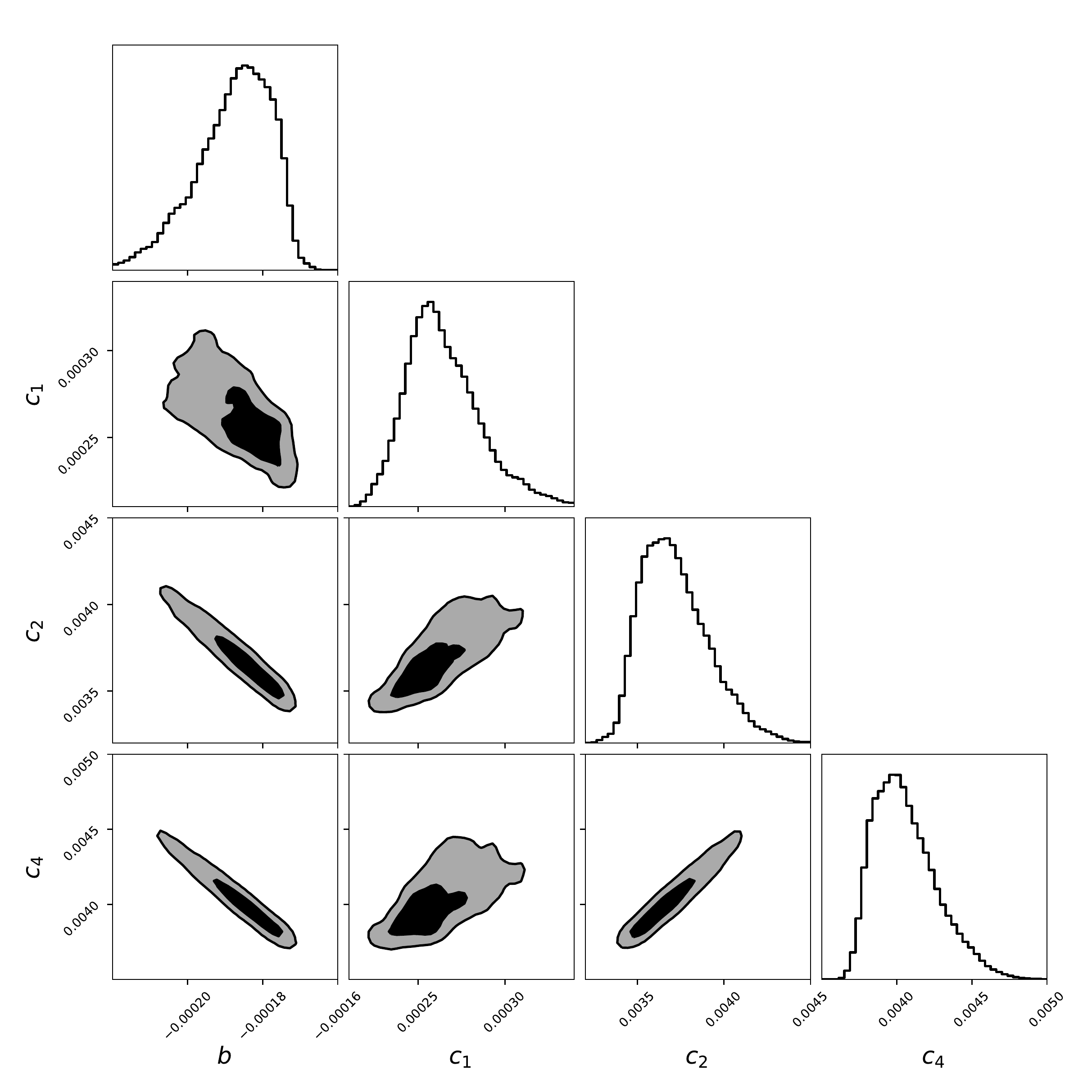}
    \\
    \includegraphics[width=0.6\textwidth]{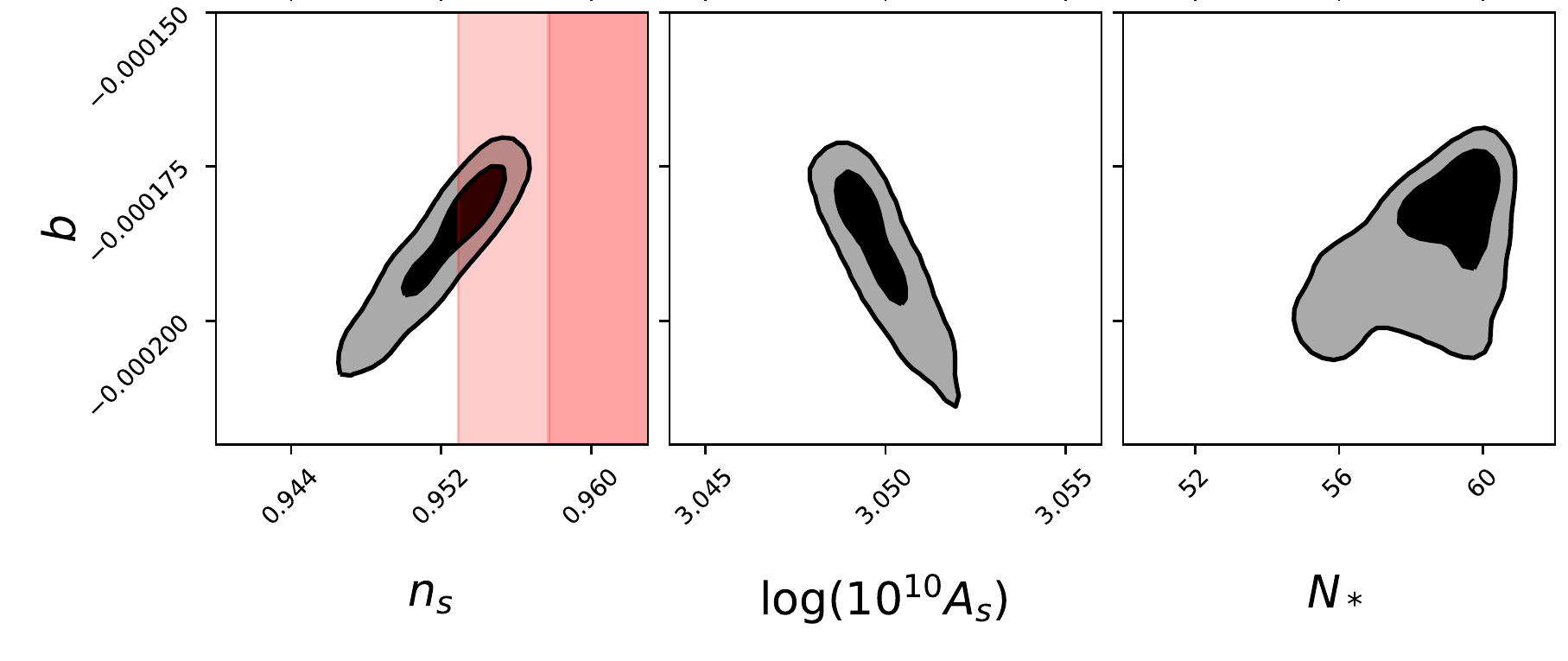}
    \caption{
    Posterior distributions in the fit of the two-field inflation model given by Eqs.~\eqref{eqn:Vtildertheta} and \eqref{eqn:bcxisymmetries} to CMB data from {\it Planck} 2018, with a prior that the model produce PBHs which could comprise all of DM.
    The last row also shows the posteriors between $b$ and derived observables, which are optimized over possible reheating histories.
    The black and grey filled contours correspond to the 68\% and 95\% deviations from the distribution means. 
    The red shaded regions show the 68\% and 95\% CL for $n_s (k_*)$ from {\it Planck} 2018.
    }
    \label{fig:MCMC_corner}
\end{figure*}
\begin{figure*}
    \includegraphics[scale=0.5]{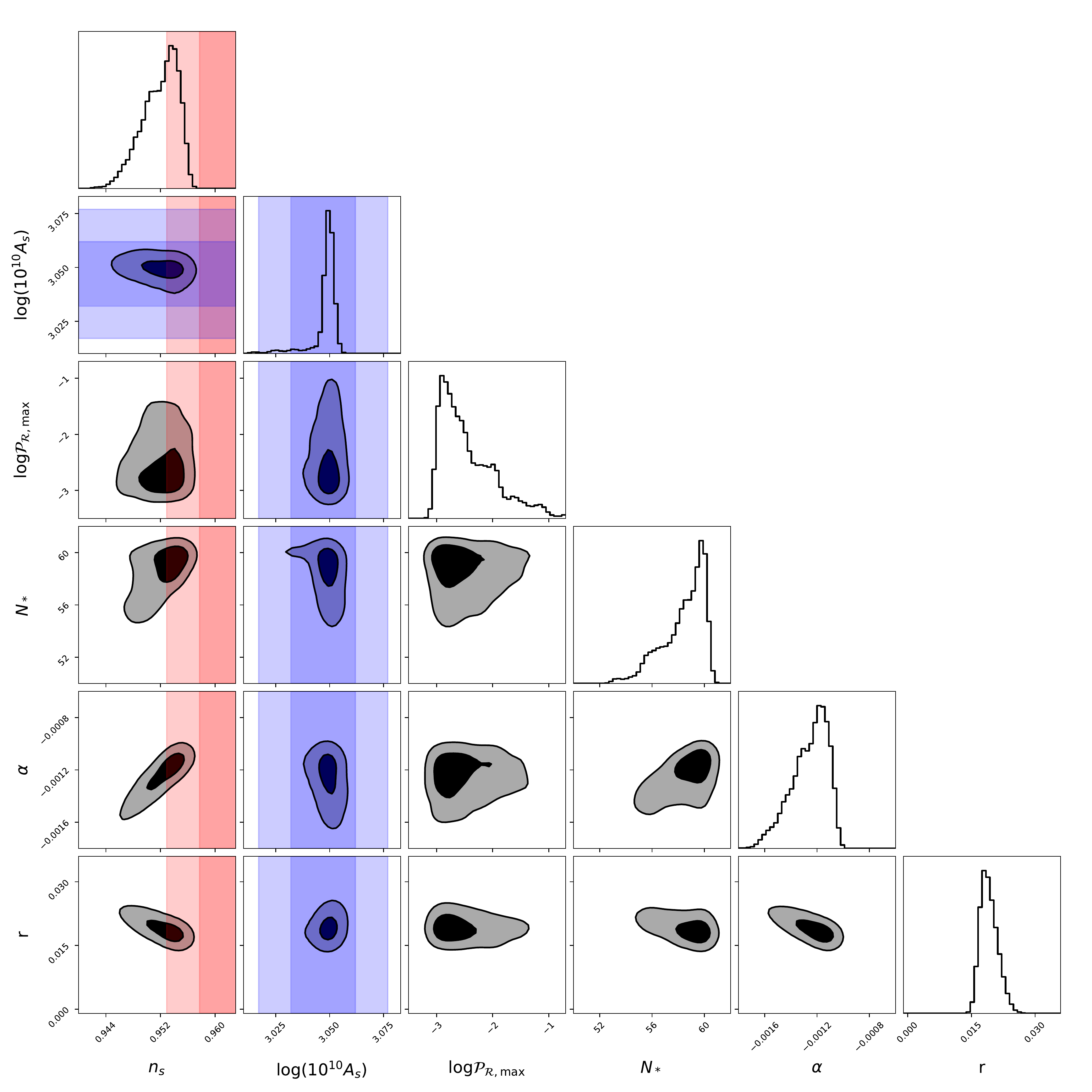}
    \caption{
    Posterior distributions on $n_s (k_*)$, $A_s (k_*)$, the peak of $\mathcal{P}_\mathcal{R}$, $N_*$, $\alpha (k_*)$, and $r(k_*)$; all of these quantities are optimized over possible reheating histories.
    The red shaded regions show the 68\% and 95\% CL for $n_s (k_*)$ from {\it Planck} 2018, and the blue shaded regions show the 68\% and 95\% CL for $A_s (k_*)$ from {\it Planck} 2018.
    There is a correlation between $n_s (k_*)$ and $N_*$. 
    At the high-end cutoff of $N_*=60$, the contours show that higher values of $n_s (k_*)$ are preferred; as $N_*$ decreases, $n_s (k_*)$ decreases as well, eventually falling well outside of the {\it Planck} 2018 constraints.
    Whereas there is also a clear correlation between $n_s (k_*)$ and $A_s (k_*)$, the variations in $A_s (k_*)$ are much less than the {\it Planck} 2018 1$\sigma$ errors, hence $A_s (k_*)$ is not a key constraint.
    }
    \label{fig:obs_corner}
\end{figure*}

The main results of this analysis are shown in Table~\ref{tab:best_fit_params} and Figs.~\ref{fig:MCMC_corner} and \ref{fig:obs_corner}.
Table~\ref{tab:best_fit_params} shows the marginalized posterior means, best-fit values, and corresponding error bars on model parameters, as well as the constraints on their ratios. The maximum likelihood model has $b=-1.73 \times 10^{-4}$, $c_1 = 2.34 \times 10^{-4}$, $c_2 = 3.42 \times 10^{-3}$, and $c_4 = 3.75 \times 10^{-3}$, and yields predictions for CMB observables $n_s (k_*) = 0.9560$, $\alpha (k_*)=-0.001$, $r (k_*)=0.016$, and $\log(10^{10} A_s)=3.048$, in excellent compliance with {\it Planck} constraints. 
This demonstrates that there is a region of parameter space in our model that is both compatible with {\it Planck} constraints and can produce PBH DM.

The posterior distributions on the model parameters are shown in Fig.~\ref{fig:MCMC_corner}, and posterior distributions for derived (cosmological) parameters are shown in Fig.~\ref{fig:obs_corner}. 
Derived parameters are analyzed in post-processing of the MCMC chains, for a subset of $\approx 4 \times 10^4$ samples. Consistent with \texttt{emcee} documentation \cite{2013PASP..125..306F}, we find the marginalized constraints on the model parameters $\{ b,c_1,c_2,c_4\}$ from this subset of steps are near-identical to those from the full MCMC chains, thus validating our use of a subset of points for constraints on derived parameters, such as $n_s$ and $A_s$.

From Fig.~\ref{fig:MCMC_corner}, we can clearly see the degeneracies discussed in \S~\ref{sec:degen1}.
As expected, $b$ is anticorrelated with the other $c_i$'s, while all the $c_i$ parameters are positively correlated with one another.
Moreover, at the larger end of the posterior distribution for $b$, we see a sharp cutoff, whereas toward smaller values there is a more gradual tail.
Due to the anticorrelation, this behavior is reversed for the $c_i$; the posteriors exhibit cutoffs at small values and a tail at larger values.

We can understand this behavior if we look at the marginalized posteriors for $b$ and the cosmological observables, a subset  of which is shown in the last row of Fig.~\ref{fig:MCMC_corner}.
The parameter $b$ shows a clear positive correlation with $n_s (k_*)$; as we move along the contour to more negative values for $b$, $n_s (k_*)$ decreases past the $2\sigma$ {\it Planck} 2018 error bars.
Hence, towards smaller values of $b$ (larger values of $c_i$), the posteriors show a tail corresponding to the Gaussian prior on the value of $n_s (k_*)$.
There is also a correlation between $b$ and $N_*$: larger values of $b$ produce models that prefer a larger $N_*$. Hence at a large enough value for $b$, we are eventually constrained by the requirement that $N_*$ be less than 60.
This explains the sharp cut-offs observed in the posteriors.

We reiterate that the distributions of the $n_s$, $A_s$, $N_*$, and the other observables shown in Figs.~\ref{fig:MCMC_corner} and \ref{fig:obs_corner} are the {\it optimal} values for a given set of model parameters. 
That is, for a given set of model parameters, we consider the optimal reheat history, rather than marginalizing over possible reheat histories. 
From Fig.~\ref{fig:obs_corner}, one may appreciate that in comparison with the {\it Planck} measurement of $A_s$, the (optimal) $A_s$ is essentially fixed in our model. Underlying this, however, is a delicate $A_s$-$n_s$ compensation in the choice of $N_*$, which in turn generates a spread of $N_*$ values (see, e.g., the $n_s-N_*$ posterior), in contrast to what one might expect if $n_s$ were solely driving the constraints on the models (which would lead to $N_*=60$). 
The optimization of reheat history is discussed further in Sec.~\ref{sec:deg}. 

In addition, although the resulting PBH masses tend to populate the lower end of the allowed DM window, $ 10^{-16}\lesssim M_{\textrm{PBH}}/M_\odot \lesssim 10^{-11}$, we find regions of parameter space that yield PBHs within that mass range. The tendency to produce lower mass black holes is driven by the {\it Planck} 2018 data; compliance with measurements at the pivot scale drives the parameters towards models with $\Delta N \lesssim 14$, which is consistent with the results of Ref.~\cite{Geller:2022nkr}. (See also Refs.~\cite{Byrnes:2018txb,Carrilho:2019oqg,Ando:2020fjm}.)
However, we emphasize that the estimation of the required $\Delta N$ range to obtain black holes in this window neglected the effects of non-Gaussianity in large-amplitude curvature perturbations, which would enhance power in large fluctuations and hence yield a higher probability of producing larger black holes~\cite{Byrnes:2012yx,Young:2013oia,Pattison:2017mbe,Biagetti:2018pjj,Kehagias:2019eil,Ezquiaga:2019ftu,Ando:2020fjm,Tada:2021zzj,Biagetti:2021eep,Ferrante:2022mui}.
We leave incorporating these effects to future studies.

Finally, while the model parameters are constrained at the $\sim 10 \%$ level, this is deceptive because of the strong degeneracies in parameter space.
If we instead examine the ratios of the parameters, as shown in the bottom rows of Table~\ref{tab:best_fit_params}, we see that these are constrained at the percent level. Hence, the parameters must be tuned to less than a percent in order for the model to produce an appropriate population of PBHs while also remaining in compliance with existing measurements.

\subsection{Degeneracy directions and compensations}
\label{sec:deg}

As mentioned in Sec.~\ref{sec:degen1}, we identify degeneracy directions in parameter space: directions along which parameter variations will leave the potential and power spectrum, and thus predictions for CMB observables, unchanged. In this section we study this quantitatively. 

\begin{figure}[h]   
    \centering
    \includegraphics[width=.5\textwidth]{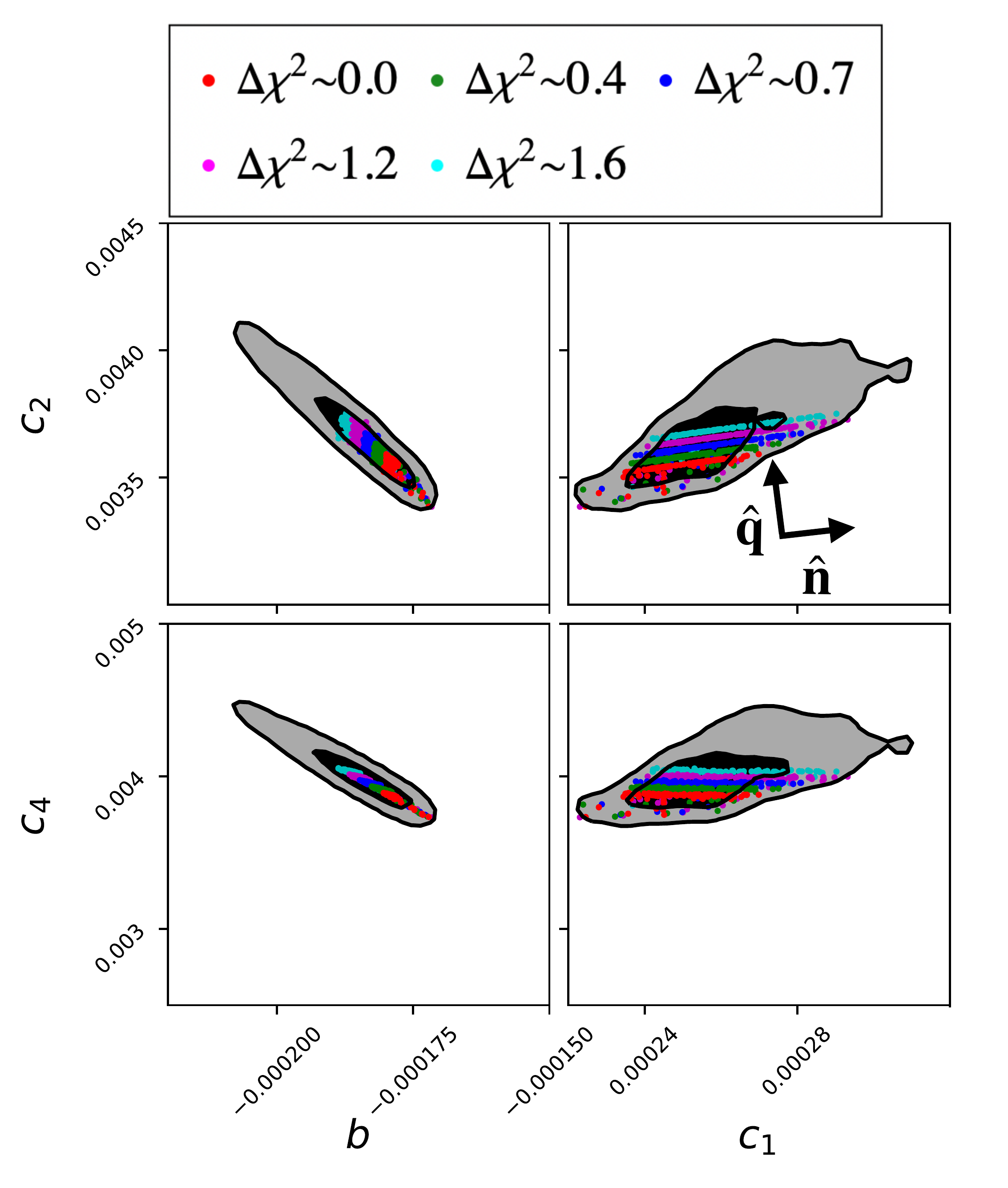}
    \caption{Four of the panels from the corner plot in Fig.~\ref{fig:MCMC_corner}, overlaid with a scatter plot of five sets of parameter sets. Points within a single color region yield nearly degenerate overall fits to the data, with differences of their total $\chi^2$ within $0.01$.
    The fiducial parameter set ${\cal F}$, for which the potential and power spectrum are plotted in Fig.~\ref{fig:V_PR_degen}, is one point within the red set of points. One example of a degeneracy direction is indicated in the top right panel by the unit vector $\hat{n}$, with orthogonal direction indicated by $\hat{q}$. Moving along $\hat{n}$ means moving along one color (with fixed $\chi^2$), whereas moving along $\hat{q}$ means moving from one color to another (changing $\chi^2$). The legend shows the approximate difference in total $\chi^2$ between each (degenerate) set of points relative to the fiducial parameter set ${\cal F}$. 
    } 
    \label{fig:bands}
\end{figure}

We consider two parameter sets to be degenerate if the difference of their total $\chi^2$ values is less than $0.01$. We define five super-sets of degenerate points (parameter sets) within the parameter space, as illustrated in the scatter plots in Fig.~\ref{fig:bands} superposed upon four of the two-dimensional posteriors, in which each super-set is assigned a color: red, green, blue, magenta, and cyan. Parameter sets within a given color yield nearly identical values of $\chi^2$; parameter sets belonging to different colors yield different values of $\chi^2$. The fiducial parameter set ${\cal F}$, used in the plots in Fig.~\ref{fig:V_PR_degen}, belongs to the set of red points in Fig.~\ref{fig:bands}.

\begin{figure}[h]   
    \centering
    \includegraphics[width=.48\textwidth]{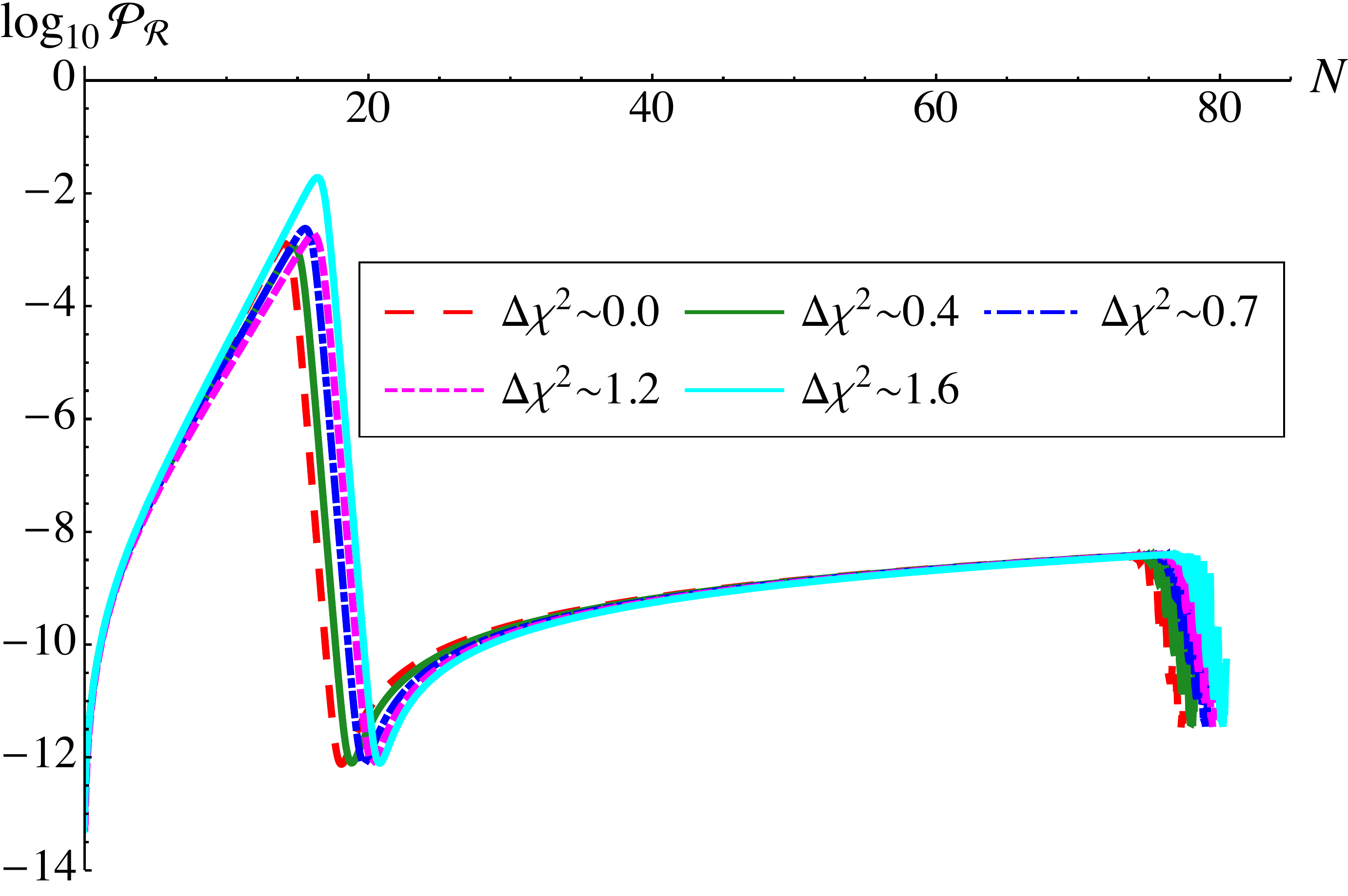}
    \caption{The power spectra for representative points in parameter space drawn from each of the five super-sets shown as distinct color bands in Fig.~\ref{fig:bands}. As in Fig.~\ref{fig:bands},
    $\Delta\chi^2$ for each parameter set is calculated relative to the $\chi^2$ of the fiducial parameter set $\mathcal{F}$ (red).}
     \label{fig:pr_rainbow}
\end{figure}

By computing the total $\Delta \chi^2$ relative to the red points, 
we can locally identify two directions within the parameter space. Along the degenerate direction $\hat{n}$, $\Delta\chi^2$ remains effectively constant (to within $0.01$). Along the orthogonal direction $\hat{q}$, $\Delta\chi^2$ changes appreciably, as shown in the legend of Fig.~\ref{fig:bands}. The power spectra for five choices of parameters are plotted in Fig.~\ref{fig:pr_rainbow}, where each curve represents a single point from the super-set of corresponding color shown in Fig.~\ref{fig:bands}. 

\begin{figure}[h]
    \includegraphics[width=.48\textwidth]{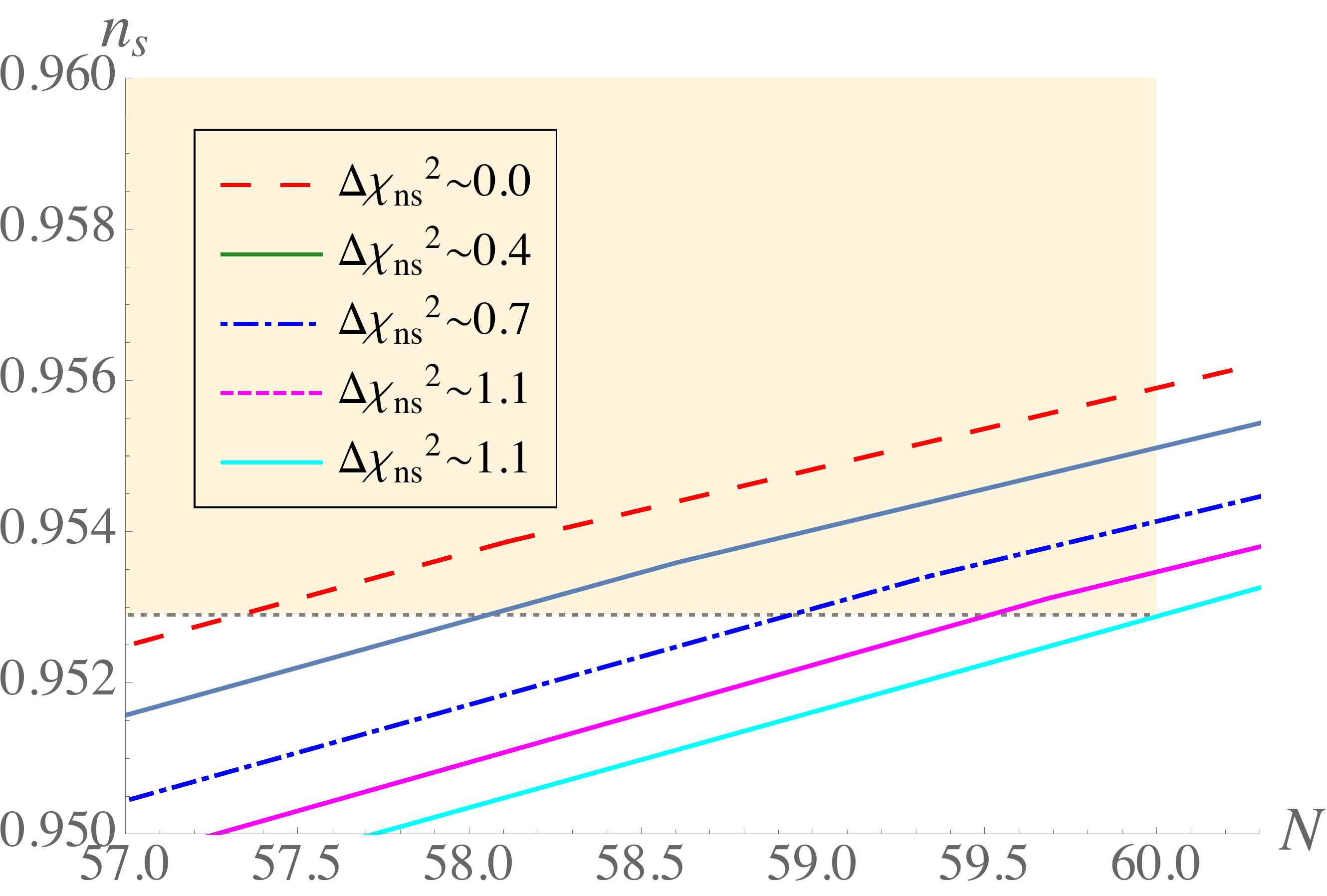}
    \caption{The spectral index $n_s$ for the same five sets of parameters used in Fig.~\ref{fig:pr_rainbow}, each drawn from the corresponding color bands in Fig.~\ref{fig:bands}.
    The legend shows differences in $\chi^2_{\rm ns}$ for each parameter set relative to the fiducial parameter set ${\cal F}$ (red). Moving along the direction $\hat{q}$ orthogonal to the degeneracy direction corresponds to moving from red to green, blue, magenta, and cyan, with increasing $\Delta \chi^2_{\rm ns}$.
    } 
    \label{fig:ns_bands}
\end{figure}

Our analysis allows $N_{*}$ to vary within a range of $(50,60)$ $e$-folds, and for a given parameter set, the ultimate fit to data is performed for the value of $N_{*}$ which minimizes $\chi^2$; we call this the \textit{optimal} $N_*$ value. From the $N_{*}-n_s$ posterior in Fig.~\ref{fig:obs_corner}, we see that a larger value of $n_s$, approaching the central {\it Planck} value, favors a smaller range of values for $N_{*}$, whereas a smaller value of $n_s$, moving away from the central {\it Planck} value, allows for a larger range of $N_{*}$. Whereas we might expect the opposite behavior due to the scaling of the running $\alpha$ with the size of $n_s$, this behavior is in fact explained by looking at the trends in $\Delta\chi^2_{n_s}$ and $\Delta\chi^2_{A_s}$ as we move from the fiducial (red) set along the orthogonal direction $\hat{q}$. Here we use the notation $\chi^2_{y_i}$ to mean the normalized $\chi^2$ given by the square of the difference between the {\it Planck} and model value divided by the $\sigma_{y_i}^2$ for observable $y_i$. 

The value of $\Delta\chi^2_{n_s}$, where the difference is calculated relative to the fiducial parameter set $\mathcal{F}$, changes as: $\Delta\chi^2_{n_s}=0$ (red), $0.4$ (green), $0.7$ (blue), $1.1$ (magenta), and $1.1$ (cyan). These values almost exactly track the overall differences $\Delta\chi^2$, for all but the cyan parameter set, showing that the change in $\Delta\chi^2$ is driven primarily by the fit to $n_s$ data. However, the optimal choice of $N_{*}$ is also somewhat driven by $A_s$. At larger values of $n_s$, i.e. closer to the \textit{Planck} central value, the compensatory behavior of $A_s$ pushes it further from its \textit{Planck} value, which results in an optimal choice of $N_*$ slightly below the value of 60 $e$-folds that would minimize $\chi^2_{n_s}$ alone. This compensation between $n_s$ and $A_s$ results in a wider spread of optimal $N_*$ values for larger $n_s$ (closer to the \textit{Planck} value) and a narrower spread for smaller $n_s$ (further from the \textit{Planck} value). At smaller $n_s$, $A_s$ is closer to the \textit{Planck} value, and $\Delta\chi^2_{A_s}$ remains small for a wider range of $N_*$ values. 

This interplay between values of $A_s$ and $n_s$ and the optimal value of $N_*$ can be seen by comparing the cyan and magenta curves in Fig.~\ref{fig:ns_bands}, which were chosen to have the same value of $n_s$ and thus equivalent $\Delta\chi^2_{n_s}$, but which have different values of $A_s$: $\Delta\chi^2_{A_s}\sim 0.4$ for cyan and $\Delta\chi^2_{A_s}\sim 0.008$ for magenta. As a result, the favored value $N_*$ is lower for the magenta parameter set (59.5) than for the cyan parameter set (60.0), indicating that the length of time for which the field experiences ultra slow-roll is longer for the cyan parameter set by about $0.5$ $e$-folds. This is consistent with the fact that the cyan parameter set shown in Figs.~\ref{fig:pr_rainbow}-\ref{fig:ns_bands} has a value of the coupling $b$ with larger magnitude $|b|$ than does the magenta parameter set. As we saw in Sec.~\ref{sec:degen1}, increasing the magnitude of $|b|$ increases the depth of the local minimum in the small-field feature of the potential, and thus lengthens the duration of ultra-slow roll. This results in a modest but noticeable increase in the 
height of the peak in $\mathcal{P}_{\mathcal{R}}$ for cyan relative to magenta, as can be seen in Fig.~\ref{fig:pr_rainbow}. The interplay between $A_s, n_s$ and the optimal choice of $N_*$ thus connects the small-field physics to the CMB-scale physics.

\section{Gravitational Wave Forecasts}

%
\begin{figure*}
    \includegraphics[scale=0.58]{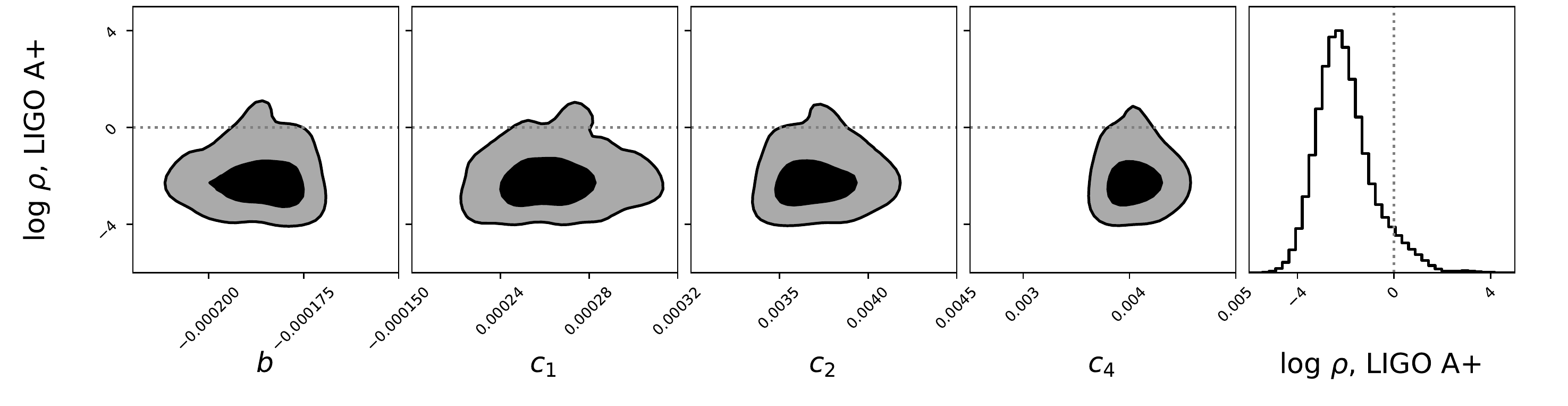}
    \includegraphics[scale=0.58]{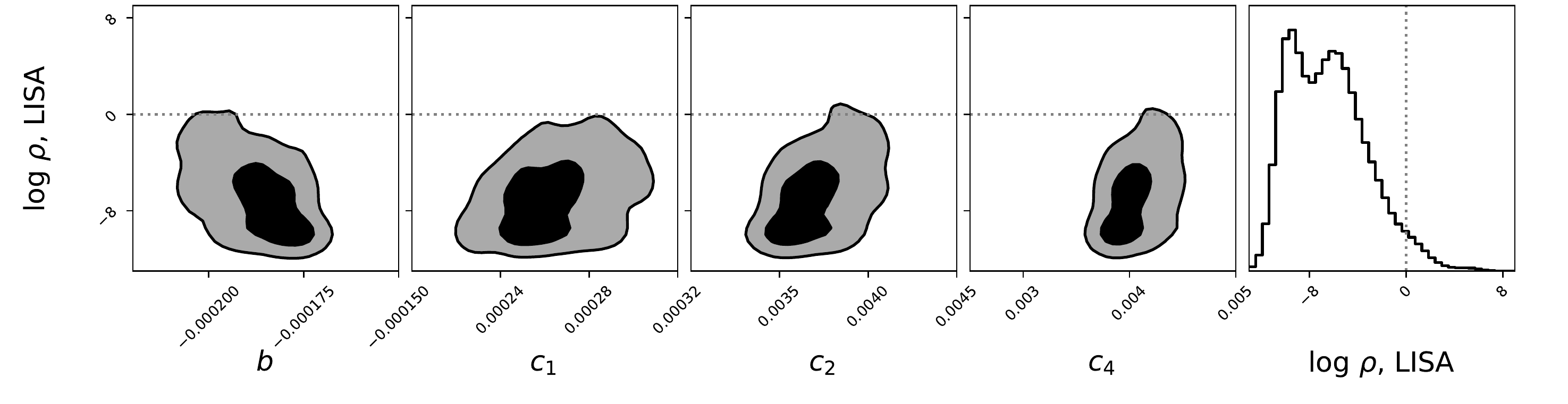}    \includegraphics[scale=0.58]{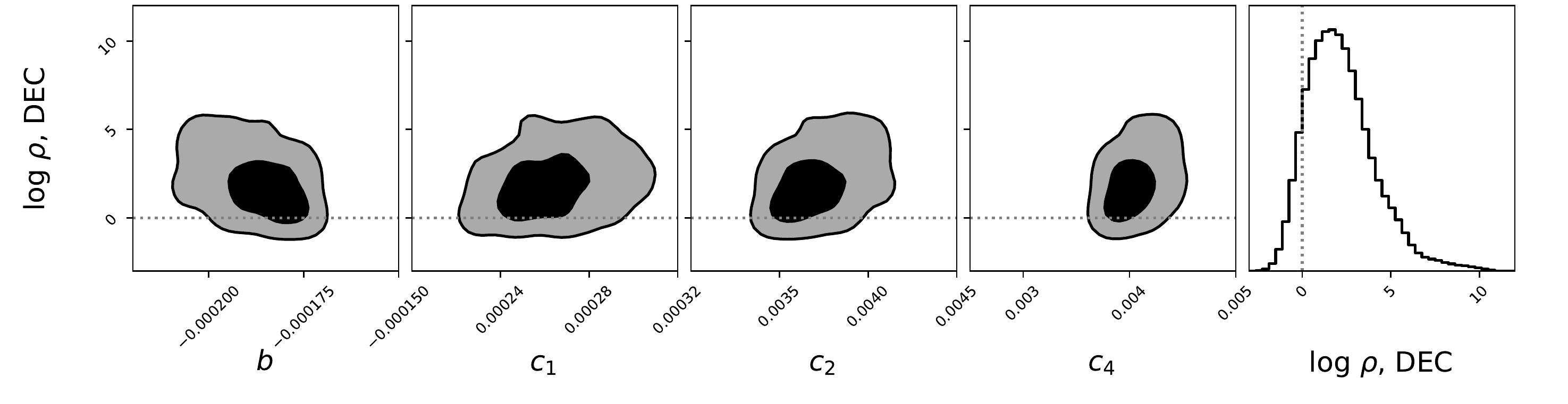}
    \includegraphics[scale=0.58]{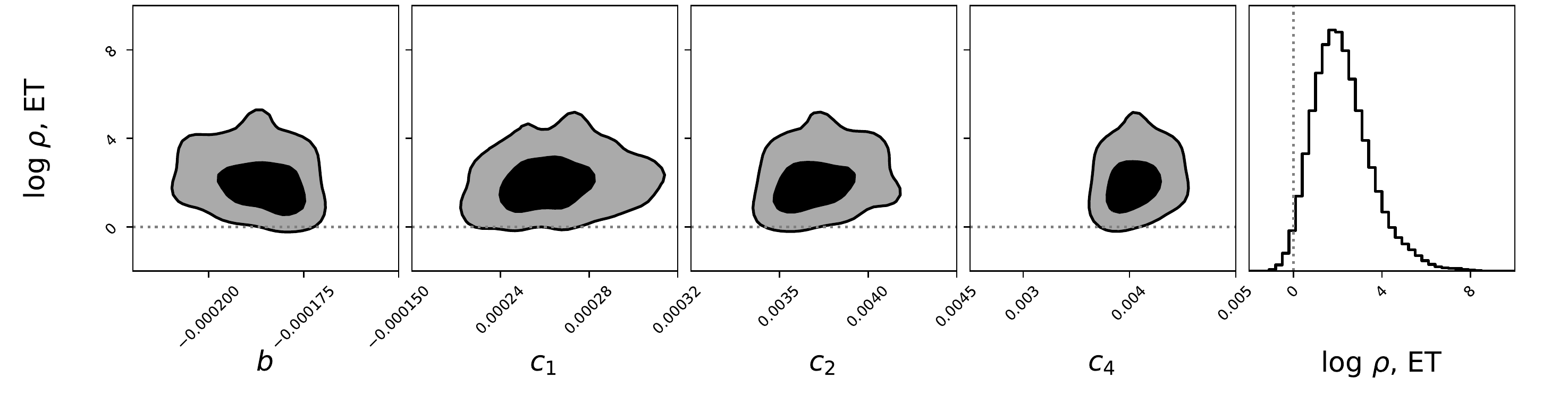}
    \includegraphics[scale=0.58]{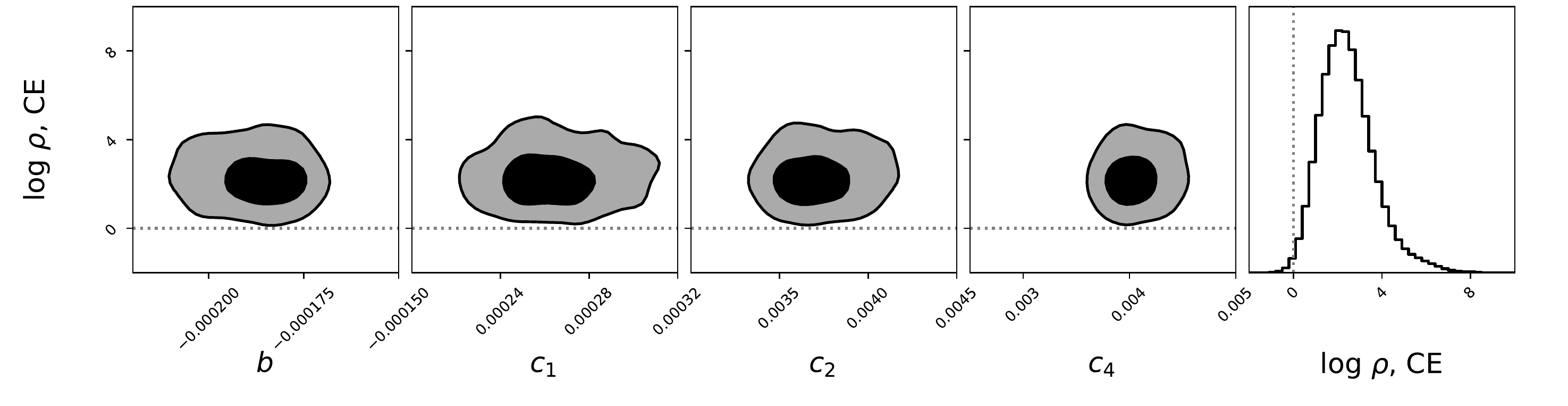}
    \caption{
    Signal-to-noise ratio ($\rho$) of gravitational waves from this multifield inflation model in LIGO A+, LISA, DECIGO, ET, and CE, shown from top to bottom respectively. Here, we use a subset of 13400 of the data points sampled by the MCMC.
    For all experiments, we find models in some parameter region with $\rho > 1$. 
    The sensitivity to these induced gravitational waves is greatest in ET and CE.
    }
    \label{fig:SNR_all}
\end{figure*}

In the previous sections, we have discussed the 68\% and 95\% CL parameter regions for our multifield inflation model in terms of the parameters $b, c_1, c_2, c_4$ and $\xi$, to obtain PBHs as all DM. 
This culminated in the marginalized constraints shown in \S~\ref{sec:results}. 
Since inflationary models with large amplifications of scalar perturbations can also produce secondary tensor perturbations, we compute the GW spectra that result from these tensor perturbations in the present universe, using the allowed parameter space in \S~\ref{sec:results}. 
We use the formulation described in \ref{sec:GWs} to compute the dimensionless present-day spectral density of GW modes contained within the Hubble radius, $\Omega_{\textrm{GW},0}h^2$, from Eq.~\eqref{eq:dimensionlessGWspectraldensity} and the corresponding SNR, $\rho$, as per Eq.~\eqref{eq:SNR} for various experiments. 
We describe in further detail the experiments and their sensitivities below.

At high frequencies, CE and ET are sensitive to the range $10$-$10^3$ Hz. 
At slightly lower frequencies, between e.g. $10^{-5}$-$10$ Hz, we also expect LISA and DECIGO to have some sensitivity to the models we consider. 
It should be noted that at lower frequencies, data from the International Pulsar Timing Array (IPTA), which comes from a combination of three pulsar timing array experiments, can search for GWs in the frequency band $10^{-9}$-$10^{-7}$ Hz. 
However, the multifield inflation models which produce PBHs that can comprise all of DM do not produce strong signals in this frequency range, hence we do not consider IPTA.

In Fig.~\ref{fig:SNR_all}, we show the SNR for the GWs from the sampled models in ET, CE, DECIGO, LISA, and LIGO A+. 
The dashed lines demarcate the threshold for observability where $\rho \geq 1$.  
We also list the mean $\rho$ for each experiment, along with 68\% and 95\% CL regions, in Table~\ref{tab:SNR}. 
ET, CE, and DECIGO are all very sensitive to the induced GWs from the sampled parameter region in Fig.~\ref{fig:MCMC_corner}, with most points yielding SNRs that are much larger than unity. 
For all three experiments, the centers of the posteriors lie at around $\rho\simeq 10^2$.
However, LISA and LIGO A+ have much less sensitivity for the favoured parameter region of Fig.~\ref{fig:MCMC_corner}. 
In the case of LISA, we see that only a very small part of the 95\% confidence region lies above $\rho=1$ and for LIGO A+, we see that there is only a slightly larger, albeit small region lying above the threshold.

The peak sensitivity for each experiment can be understood in terms of the comoving wavenumbers that exit the Hubble radius during inflation, as well as the GW frequencies. 
For ET and CE, the peak sensitivity is around $k_{\rm p}/k_{\rm eq}\simeq 10^{18}$, where $k_p$ is the position of the hypothetical peak of the primordial power spectrum and $k_{\rm eq}=0.01\,\textrm{Mpc}^{-1}$ is the comoving wavenumber that corresponds matter-radiation equality~\cite{Planck:2018vyg}.
For DECIGO the peak lies at around $k_{\rm p}/k_{\rm eq}\simeq 10^{16}$ and for LISA it is smaller still, at around $k_{\rm p}/k_{\rm eq}\simeq 10^{14}$. 
In our case, since producing PBHs as DM is an important physical requirement, we demand $k_{\rm p}=k_{\rm pbh}$, which we recall from \S~\ref{sec:PBHconstraints}. 
For the resulting marginalized posterior distribution favoured for this model, we see that the  $10^{16}\lesssim k_{\rm p}/k_{\rm eq}\lesssim 10^{18}$ is preferred, culminating in GWs favoured with frequencies lying in the mHz to kHz range; this is optimally situated for observation by ET, CE, and DECIGO. It should be noted that a very small region of the GW spectrum predicted by the posterior distribution is in the LIGO band. It has been shown that Gaussian density perturbations producing secondary GWs in the LIGO band produce PBHs that would have evaporated \cite{Kapadia:2020pnr}. However, since there is an underestimation of $\Delta N$ as described at the end of \S~\ref{sec:PBHconstraints} due to simplified assumptions, we confirm that the PBH mass would surpass the evaporation bound when propagating these effects. We intend to perform a detailed study of the PBH mass spectrum of which we leave to future work.

\begin{table}[htb!]
  \renewcommand{\arraystretch}{1.5}
  \begin{tabular}{|l|c|c|c|}
    \hline\hline 
    Experiment & $\log_{10} \, \rho$ & $68\%$ CL & $95\%$ CL \\
    \hline \hline
    LIGO A+ & $-2.25$ & ${}^{+1.18}_{-0.74}$ & ${}_{-1.34}^{+3.02}$ \\

    LISA & $-6.52$ & ${}_{-3.33}^{+3.28}$ & ${}_{-3.47}^{+6.99} $\\

    ET & 2.04 & ${}_{-0.95}^{+1.35}$ & ${}_{-1.58}^{+3.23}$ \\

    DECIGO & 1.91 & ${}_{-1.49}^{+1.91} $&$ {}_{-2.22}^{+4.49}$ \\

    CE & 2.32 & ${}_{-0.99}^{+1.17}$ & ${}_{-1.55}^{+3.15}$ \\

    \hline
  \end{tabular} 
  \caption{
    The signal-to-noise ratio ($\rho$) for relevant experiments with upper and lower $68\%$ and $95\%$CL bounds. 
    The central values and bounds are shown in log scale.
    }
  \label{tab:SNR}
\end{table}

In summary, we find tantalizing GW phenomenology predicted for the posterior distribution of model parameters in our multifield inflation model. 
The correspondence between PBHs as DM and primordial GWs remains a compelling prospect for future ground- and space-based experiments.

\section{Conclusions}
\label{sec:conclusion}

We have performed an MCMC analysis of a simple yet generic multifield inflation model characterized by two fields coupled to each other and nonminimally coupled to gravity. This model was fit to {\it Planck} 2018 data, parametrized by measurements of the spectral parameters $A_s$, $n_s$, $\alpha$, and $r$, and with a prior that the primordial power spectrum should lead to PBH production in the ultralight asteroid-mass range, where constraints still allow for PBHs to account for all of DM. We find a nontrivial region of parameter space in our model that is both compatible with {\it Planck} data and can produce PBH DM. The constraints on allowable regions of parameter space are driven in particular by $n_s$ and $N_*$.

There are a number of reasons why we choose to focus on this multifield model.
The Standard Model includes multiple scalar fields, and extensions to the Standard Model typically feature many more. Therefore, embeddings of inflationary dynamics within realistic models of high-energy physics are likely to involve multiple interacting scalar fields. Given that several types of
single-field inflation models have successfully yielded predictions for PBHs as DM while remaining in compliance with CMB data~\cite{Carr:2020gox,Carr:2020xqk,Green:2020jor,Escriva:2022duf,Escriva:2022duf,Ozsoy:2023ryl}, it seems natural to study whether multifield models can accomplish the same, while also reducing the necessary amount of fine-tuning to produce black holes. The family of models we consider is a natural generalization of multifield models that have been studied extensively in the literature~\cite{Kaiser:2012ak,Kaiser:2013sna,Schutz:2013fua,Kaiser:2015usz}, and is closely related to well-known examples such as Higgs inflation \cite{Bezrukov:2007ep,Greenwood:2012aj,Kawai:2014gqa,Kawai:2015ryj} and $\alpha$-attractors \cite{Kallosh:2013hoa,Kallosh:2013maa,Galante:2014ifa}. 

The results of our MCMC show that there is a robust region of parameter space for which this family of models can produce PBHs in the appropriate mass range to comprise all of DM, while also remaining in compliance with {\it Planck} data.
In particular, the posteriors on all parameters show a Gaussian-like tail at one end that is controlled by the measurement of $n_s (k_*)$, and a sharp cutoff at the other end from our requirement that $N_*$ remain in the range $55 \pm 5$.
Due to our procedure for optimizing over possible reheat scenarios, we find that the optimal $N_*$ typically fixes $A_s$ to the central value of the {\it Planck} measurement.

Through this analysis, we found that whereas the parameters of the model are constrained at around a 10\% level, there is a degeneracy direction in the parameter space that leads to fine-tuned ratios at the percent level. 
It is possible that the actual required level of fine-tuning is  
greater than this, given that studies of single-field inflation models have found that model parameters typically need to be fine-tuned to as much as one part in $10^7$ in order to give rise to enhancements to the power spectrum that are sufficiently large to produce PBH DM~\cite{Garcia-Bellido:2017mdw,Ezquiaga:2017fvi,Kannike:2017bxn,Germani:2017bcs,Motohashi:2017kbs,Di:2017ndc,Ballesteros:2017fsr,Pattison:2017mbe,Passaglia:2018ixg,Byrnes:2018txb,Biagetti:2018pjj,Carrilho:2019oqg,Inomata:2021tpx,Inomata:2021uqj,Pattison:2021oen}. 
On the other hand, our results suggest that our multifield model may require less fine-tuning than some well-studied single-field models.

Upon fixing $\xi$, predictions from our multifield model depend on only four free parameters. Moreover, we can shift any one of these parameters at the 10\% level, and the constraints on the parameter ratios require the remaining three to be tuned at the 1\% level, yielding a total degree of fine-tuning of approximately $10^{-7}$. However, since the most constraining quantity is $n_s$, which has error bars at the 1\% level, the relative amount of fine-tuning needed to produce PBHs is $10^{-5}$. While there exist more rigorous measures of fine-tuning~\cite{Athron:2007ry,Fowlie:2014xha}, we leave such a quantitative analysis to future work.

Furthermore, the allowed parameter region in this model produces observable GW signals in frequency ranges that future experiments such as LIGO A+ and Virgo, ET, CE, DECIGO and LISA are projected to be sensitive to. 
The observational prospects for DECIGO, ET, and CE are particularly good for this model; the latter two experiments have a central region of the MCMC posterior distribution with signal-to-noise ratio of $\rho > 100$. 
This result suggests that this inflation model is a viable and well-motivated candidate to explain both the observed DM and to generate observationally relevant primordial GWs.
 
Given these results, there are a number of interesting directions for future work. First, since we find that the spectral index $n_s$ is especially constraining for these models, it may be instructive to look at forecasts for constraints by CMB-S4~\cite{Abazajian:2019eic,CMB-S4:2016ple}. Moreover, improving measurements on the running of the spectral index $\alpha (k_*)$ could also play an important role in helping to distinguish among such models. In addition, predictions for the tensor-to-scalar ratio, $r (k_*) \simeq 0.016$, are approximately a factor of two below current observational bounds \cite{BICEP:2021xfz}, whereas similar models (with multiple interacting scalar fields and nonminimal couplings, but with no small-field features that could yield PBHs) tend to predict considerably smaller values, $r (k_*) \simeq 0.004$ \cite{Kaiser:2013sna}. Thus we expect measurements of $r(k_*) $ to play a key role in future tests of this model. Along these same lines, it would be interesting to consider multifield models that combine the suppression of $r(k_*)$ found in Ref.~\cite{McDonough:2020gmn} with the enhancement on small scales studied here. We leave this for future work.

Second, in this work, we examined the SGWB resulting from inflation models that can produce PBHs as DM. It would also be interesting to perform a dedicated MCMC analysis to find the regions of parameter space for such models that produce detectable GWs, regardless of the implications for DM. 

Third, given the discussion above of the amount of fine-tuning in our model, another direction one could pursue would be a Bayesian comparison of our model versus similar single-field models, or to further consider more rigorous quantitative measures of fine-tuning in these classes of models. (See, e.g., Ref.~\cite{Cole:2023wyx}.)

Finally, a more systematic analysis of uncertainties associated with the post-inflation reheating phase \cite{Martin:2021frd} would be valuable for considering how the domain of viable parameter space for the multifield models studied here compares with those of other types of inflationary models. It is likely that if we were to sample $N_*$ within the range $55 \pm 5$ and then marginalize over it, that would increase the error bars in the estimates of our parameter constraints, which in turn would reduce the implied degree of fine-tuning for such models. Such questions remain a topic for further research.

\section*{Acknowledgements}

We are pleased to thank Christian Byrnes, Philippa Cole, Alan Guth, Mikhail Ivanov, Benjamin Lehmann, J\'{e}r\^{o}me Martin, Subodh Patil, Salvatore Vitale, and Vincent Vennin for helpful discussions. This material is based upon work supported by the U.S. Department of Energy, Office of Science, Office of High Energy Physics of U.S. Department of Energy under grant Contract Number DE-SC0012567.  
W.Q. was additionally supported by National Science Foundation Graduate Research Fellowship under Grant Nos. 1745302 and 2141064. 
S.B. is supported by funding from the European Union's Horizon 2020 research and innovation programme under grant agreement No.~101002846 (ERC CoG ``CosmoChart'') as well as support from the Initiative Physique des Infinis (IPI), a research training program of the Idex SUPER at Sorbonne Universite. E.M. is supported in part by a Discovery Grant from National Sciences and Engineering Research Council of Canada.

\begin{widetext}
\appendix

\section{The power spectrum of scalar curvature perturbations during ultra-slow-roll} 
\label{app:USR}

We may expand the spacetime line element to linear order in scalar metric perturbations around a spatially flat FLRW metric as in Ref.~\cite{Bassett:2005xm}
\beq
ds^2 = - (1 + 2 A) dt^2 + 2 a \left( \partial_i B \right) dt \, dx^i + a^2 \left[ \left( 1 - 2 \psi \right) \delta_{ij} + 2 \partial_i \partial_j E \right] dx^i dx^j .
\label{dsFLRWscalar}
\eeq
The metric functions $A, B, \psi$, and $E$ are each gauge dependent, as are the field fluctuations $\delta \phi^I$ identified in Eq.~(\ref{phivarphi}). We may construct the gauge-invariant curvature perturbation ${\cal R}$ as the linear combination \cite{Bassett:2005xm}
\beq
{\cal R} \equiv \psi - \frac{ H}{\rho + p} \delta q ,
\label{Rdefpsi}
\eeq
where $\rho = \frac{1}{2} \dot{\sigma}^2 + V$, $p = \frac{1}{2} \dot{\sigma}^2 - V$, and
\beq
\delta q = - {\cal G}_{IJ} \dot{\varphi}^I \delta \phi^J = - \dot{\sigma} \hat{\sigma}_J \delta \phi^J ,
\label{deltaq}
\eeq
with $\dot{\sigma}$ and $\hat{\sigma}^I$ defined in Eqs.~(\ref{dotsigmadef})--(\ref{hatsigma}). In a multifield model we may project onto the hypersurface of the field-space manifold that is orthogonal to the direction of the background fields' motion via \cite{Kaiser:2012ak}
\beq
\hat{s}^{IJ} \equiv {\cal G}^{IJ} - \hat{\sigma}^I \hat{\sigma}^J ,
\label{hatsIJ}
\eeq
in terms of which we may define the ${\cal N} - 1$ remaining gauge-invariant scalar perturbations
\beq
\delta s^I \equiv \hat{s}^I_{\>\> J} \delta \phi^J .
\label{sIdef}
\eeq
Although the field-space vector $\delta s^I$ includes ${\cal N}$ components, only ${\cal N} - 1$ are linearly independent. Moreover, where the field fluctuations $\delta \phi^I$ are gauge dependent, the perturbations $\delta s^I$ are gauge independent, up to linear order in fluctuations \cite{Kaiser:2012ak}. 

In terms of these quantities, the derivative with respect to cosmic time $t$ of a mode ${\cal R}_k (t)$ is given by \cite{Kaiser:2012ak}
\beq
\dot{\cal R}_k = \frac{ 2 H}{\dot{\sigma}} \left( \omega_J \delta s_k^J \right) + \frac{ H}{\dot{H}} \frac{ k^2}{a^2} \Psi_k .
\label{dotR}
\eeq
Here $\omega^I$ is the covariant turn-rate vector defined in Eq.~(\ref{omegaturnratedef}) and $\Psi$ is the gauge-invariant Bardeen potential \cite{Bassett:2005xm}
\beq
\Psi \equiv \psi + a^2 H \left( \dot{E} - \frac{B}{a} \right) .
\label{PsidefBardeen}
\eeq
Upon using Eq.~(\ref{eombackground}) for $\dot{H}$ and Eq.~(\ref{eqn:slowroll}) for $\epsilon$, Eq.~(\ref{dotR}) is equivalent to
\beq
\frac{ \dot{\cal R}_k}{H} = \frac{1}{ M_{\rm pl} H} \sqrt{ \frac{2}{\epsilon}} \left( \omega_J \delta s_k^J \right) - \frac{1}{ \epsilon} \left(\frac{ k}{aH} \right)^2 \Psi_k .
\label{dotR2}
\eeq
During ordinary slow-roll evolution of the background fields $\varphi^I (t)$, with $\epsilon \sim 10^{-3} - 10^{-2}$, we find the usual result that modes ${\cal R}_k (t)$ will remain conserved outside the Hubble radius, $\dot{\cal R}_k \simeq 0$, in effectively single-field models, for which $\vert \omega_J \delta s_k^J \vert / (M_{\rm pl} H) \sim 0$. Yet even in such (effectively) single-field scenarios, modes ${\cal R}_k (t)$ may undergo significant growth outside the Hubble radius during a phase of ultra-slow-roll evolution, as $\epsilon$ falls exponentially, $\epsilon (t_{\rm usr}) \ll 1$. (See, e.g., Refs.~\cite{Kinney:2005vj,Martin:2012pe}.)

In the limit $\vert \omega_J \delta s^J \vert / (M_{\rm pl} H) \sim 0$, the equation of motion for modes ${\cal R}_k$ assumes the simple form \cite{Bassett:2005xm}
\beq
\frac{1}{ a^3 \epsilon} \frac{d}{dt} \left( a^3 \epsilon \dot{\cal R}_k \right) + \frac{ k^2}{a^2} {\cal R}_k = 0 .
\label{eomR}
\eeq
Using the definitions of the slow-roll parameters $\epsilon$ and $\eta$ in Eq.~(\ref{eqn:slowroll}), this is equivalent to
\beq
\ddot{\cal R}_k + 3 H \left( 1 + \frac{4}{3} \epsilon - \frac{2}{3} \eta \right) \dot{\cal R}_k + \frac{ k^2}{a^2} {\cal R}_k = 0 .
\label{eomR2}
\eeq
During ultra-slow-roll, the slow-roll parameter $\epsilon$ falls rapidly, $\epsilon \rightarrow 0^+$. In that limit, the amplitude of modes ${\cal R}_k$ that have crossed outside the Hubble radius (with $k \ll aH$) will grow whenever $\eta > 3/2$ \cite{Kinney:2005vj,Martin:2012pe,Byrnes:2018txb,Liu:2020oqe,Cole:2022xqc}. Near an inflection point of the potential, for which $V_{, \sigma} \simeq 0$, the background equations of motion yield $\epsilon \sim a^{-6}$ and hence $\eta \rightarrow  3$. For small-field features in the potential of the sort we analyze here, however, $V_{, \sigma}$ need not vanish identically during the entire duration of ultra-slow-roll, and $\eta$ can exceed 3. More generally, $\eta (t)$ need not remain constant during the ultra-slow-roll phase.

Given that the amplitude of modes can grow outside the Hubble radius while $\eta > 3/2$, departures from the typical slow-roll evolution are characterized by the quantity
\beq
{\cal U} \equiv \int_{t_s}^{t_e} dt \left( \eta(t) - \frac{3}{2} \right) H (t) = \left( \bar{\eta} - \frac{3}{2} \right) N_{\rm usr} ,
\label{Udef}
\eeq
where $t_s$ and $t_e$ indicate the start and end of the ultra-slow-roll phase, each determined by the times when $\eta$ crosses $3/2$, and $\bar{\eta}$ denotes the average of $\eta (t)$ between $t_s$ and $t_e$. Given that $\epsilon \rightarrow 0^+$ during ultra-slow-roll, $\dot{H} \sim 0$ and the duration of ultra-slow-roll (in $e$-folds) may be approximated as $N_{\rm usr} \simeq H (t_e - t_s)$. 

The rapid fall of $\epsilon$ after the start of ultra-slow-roll means that some modes with wavenumber $k < k_s$ will become amplified after crossing outside the Hubble radius, beginning with wavenumber $k_{\rm min}$ given by \cite{Liu:2020oqe}
\beq
k_{\rm min} = k_s \exp \left[ - {\cal U} \right] \simeq k_s e^{- (\bar{\eta} - 3/2 ) N_{\rm usr} } ,
\label{kmin}
\eeq
where $k_s = a (t_s) H (t_s)$ is the wavenumber of the mode that first exits the Hubble radius at the start of ultra-slow-roll. The value of $k_{\rm min}$ comes from balancing the fall in amplitude of the ``decaying" contribution to ${\cal R}_k$ between the time that mode crosses outside the Hubble radius ($t_k$) and the onset of ultra-slow-roll ($t_s$), with the growth of that same term during the ultra-slow-roll phase ($t_s \leq t \leq t_e$) \cite{Liu:2020oqe}.

For modes with $k_{\rm min} \leq k \leq k_s$, which cross outside the Hubble radius prior to the onset of ultra-slow-roll, the power spectrum will be modified compared to the slow-roll expression ${\cal P}_{\cal R}^{\rm SR} (k)$ of Eq.~(\ref{eqn:PRHepsilon}) as \cite{Liu:2020oqe} (see also Refs.~\cite{Martin:2012pe,Byrnes:2018txb,Cole:2022xqc})
\beq
{\cal P}_{\cal R} (k) \simeq \left( \frac{ k}{k_s} \right)^4 \exp\left[ 4 \, {\cal U} \right] \, {\cal P}_{\cal R}^{\rm SR} (k ) 
\simeq \left( \frac{ k}{k_s} \right)^4 {\cal P}_{\cal R}^{\rm SR} (k)\, e^{( 4 \bar{\eta} - 6 ) N_{\rm usr}} \quad {\rm for} \> \, k_{\rm min} \leq k \leq k_s .
\label{PRusrkmin}
\eeq
The steep growth $\sim k^4$ begins at $k_{\rm min}$ and reaches a peak at $k_s$. For modes that exit the Hubble radius during ultra-slow-roll, with $k_e \leq k \leq k_s$, the power spectrum falls from its peak at ${\cal P}_{\cal R} (k_s)$, since those modes merely oscillate prior to exiting the Hubble radius and then experience a shorter duration ($N < N_{\rm usr}$) of growth outside the Hubble radius during the remainder of the ultra-slow-roll phase. For modes that cross outside the Hubble radius later than $t_e$, the background system again undergoes ordinary slow-roll evolution, so the standard expression for ${\cal P}_{\cal R}^{\rm SR} (k)$ of Eq.~(\ref{eqn:PRHepsilon}) applies for $k > k_e$ \cite{Martin:2012pe,Liu:2020oqe,Byrnes:2018txb,Cole:2022xqc}.

Incorporating the added growth of various modes during ultra-slow-roll, the dimensionless power spectrum reaches its maximum value at $k_s$ \cite{Martin:2012pe,Liu:2020oqe,Byrnes:2018txb,Cole:2022xqc}:
\beq
{\cal P}_{\cal R, \rm max} (k)  = {\cal P}_{\cal R} (k_s) \simeq \exp \left[ 4 \, {\cal U} \right] {\cal P}_{\cal R}^{\rm SR} (k_s)  .
\label{PRmaxUSR}
\eeq
The usual slow-roll expression for ${\cal P}_{\cal R}^{\rm SR} (k)$ in Eq.~(\ref{eqn:PRHepsilon}) already incorporates parametric growth comparable to the term $\exp [ 4 \, {\cal U} ]$, given that ${\cal P}_{\cal R}^{\rm SR} (k) \propto 1 / \epsilon$ and $\epsilon \sim a^{-6} \sim \exp [6 N_{\rm usr}]$ during ultra-slow-roll. This amplification matches the ultra-slow-roll amplification factor $\exp [ 4 \, {\cal U} ]$ if $\bar{\eta} = 3$ during ultra-slow-roll. The main impacts of the ultra-slow-roll phase, compared to the usual slow-roll expression, are then twofold: the peak value ${\cal P}_{\cal R, \rm max}$ can exceed the value represented by ${\cal P}_{\cal R, \rm max}^{\rm SR}$ if $4 \, {\cal U} > 6 N_{\rm usr}$; and the wavenumber corresponding to the peak in ${\cal P}_{\cal R} (k)$ shifts to shorter values compared to ${\cal P}_{\cal R}^{\rm SR} (k)$, due to the added growth of some modes after they cross outside the Hubble radius. In particular, whereas ${\cal P}_{\cal R}^{\rm SR} (k)$ typically reaches its {\it minimum} value at $k_s$ and its maximum value at $k_e$, the full expression for ${\cal P}_{\cal R}$ reaches its {\it maximum} value at $k_s \simeq k_e \exp[ - N_{\rm usr}]$ \cite{Liu:2020oqe}.

We may consider the impact of the ultra-slow-roll phase on ${\cal P}_{\cal R} (k)$ for our family of models in typical regions of parameter space. We select a point among the red regions of Fig.~\ref{fig:bands}, near our fiducial parameter set. For this set of parameters, the system enters ultra-slow-roll evolution (with $\eta > 3/2$) at $N_s = 18.12$ 
$e$-folds before the end of inflation and exits ultra-slow-roll at $N_e = 14.59$, 
for a total duration $N_{\rm usr} = 3.53$ $e$-folds of ultra-slow-roll. 

If we neglect the effects of ultra-slow-roll on the power spectrum, then the slow-roll approximation to the power spectrum reaches a peak value ${\cal P}_{\cal R, \rm max}^{\rm SR} = {\cal P}_{\cal R}^{\rm SR} (k_e) = 1.18 \times 10^{-3}$ at $N_e$, consistent with the behavior of most regions of parameter space under study here (see the posterior distribution of ${\cal P}_{\cal R, \rm max}$ in Fig.~\ref{fig:obs_corner}). Given ${\cal P}_{\cal R}^{\rm SR} (k_s) = 7.14 \times 10^{-13}$, this means that the slow-roll approximation to the power spectrum grows by a factor $1.66 \times 10^9$ during the ultra-slow-roll phase, consistent with the parametric growth noted above: ${\cal P}_{\cal R}^{\rm SR} (k) \propto 1 / \epsilon \sim \exp [ 6 N_{\rm usr} ]$. For these parameters, meanwhile, the ultra-slow-roll amplification factor ${\cal U}$ defined in Eq.~(\ref{Udef}) is ${\cal U} = 5.70$. We thus find $4 \, {\cal U} = 22.80 > 6 N_{\rm usr} = 21.18$, which indicates that ${\cal P}_{\cal R, \rm max} > {\cal P}_{\cal R, \rm max}^{\rm SR}$. In particular, the additional growth during the ultra-slow-roll phase yields ${\cal P}_{\cal R, \rm max} = \exp \left[ 4 \, {\cal U} \right] {\cal P}_{\cal R}^{\rm SR} (k_s) = 5.70 \times 10^{-3}$, a factor $\sim 5$ greater than the peak predicted by the slow-roll approximate form ${\cal P}_{\cal R}^{\rm SR} (k)$. 

Although details of the impact of ultra-slow-roll on ${\cal P}_{\cal R} (k)$ depend on the model and parameter set under consideration, the growth of ${\cal P}_{\cal R, \rm max}$ compared to ${\cal P}_{\cal R, \rm max}^{\rm SR}$ that we find here is consistent with previous studies, which typically find ${\cal P}_{\cal R, \rm max} \sim {\cal O} (10) \times {\cal P}_{\cal R, \rm max}^{\rm SR}$ \cite{Kinney:2005vj,Martin:2012pe,Byrnes:2018txb,Liu:2020oqe,Cole:2022xqc}. Given that the majority of points in parameter space that we sampled in our MCMC analysis yield ${\cal P}_{\cal R, \rm max}^{\rm SR} \sim 10^{-3}$, we therefore conclude that additional growth from the ultra-slow-roll phase in this family of models is consistent with ${\cal P}_{\cal R, \rm max} \lesssim 10^{-2}$ across most regions of parameter space, and hence should evade constraints (not incorporated here) based on overproducing PBHs, producing excessive spectral $\mu$-distortions, and similar small-scale effects. (Evaluating such constraints typically requires moving beyond the approximation of a Gaussian probability distribution function for the scalar perturbations, and hence remains beyond the scope of our present analysis. See, e.g., Refs.~\cite{Gow:2020bzo,Gow:2022jfb}.)

\section{The power spectrum of induced tensor perturbations}
\label{app:tensor_spectrum}

We provide the complete equations for the computation of the tensor power spectrum $P_h$ used in Eq.~\eqref{eq:GWspectrum}. 
Let us begin with the definition of linear perturbations in the conformal Newtonian gauge for the metric of the form
\begin{align}
ds^2 = -a^2(\tau)(1+2\Phi) d\tau^2 +  a^2(\tau) \left[(1-2\Psi)\delta_{ij} + \frac{1}{2} h_{ij} \right] dx^i dx^j, 
\end{align}
where $\Phi = A - \partial_t [ a^2 (\dot{E} - B / a) ]$ is the Newtonian potential and $\Psi$ is the Bardeen curvature potential defined in Eq.~(\ref{PsidefBardeen}), while $A, B$, and $E$ are defined in Eq.~(\ref{dsFLRWscalar}). We define
\begin{align}
h_{ij}(\vec{x},\tau) = \int \frac{d^3k}{(2\pi)^{3/2} } [e_{ij}^+(\vec{k})h_+(\vec{k},\tau) + e_{ij}^\times (\vec{k})h_\times(\vec{k},\tau)] e^{i\vec{k}\cdot\vec{x}},
\end{align}
which is the linear tensor perturbation including the two polarization modes. The transverse-traceless polarization tensors are
\begin{align}
    e_{ij}^+(\vec{k}) &= \frac{[e_i^1(\vec{k})e_j^1(\vec{k}) - e_i^2(\vec{k})e_j^2(\vec{k})]}{\sqrt{2}},& 
e_{ij}^\times(\vec{k}) = \frac{[e^1_i(\vec{k}) e^2_j(\vec{k}) + e^2_i(\vec{k}) e^1_j(\vec{k})]}{\sqrt{2}},
\end{align}
which are expressed in terms of orthonormal basis vectors $\textbf{e}^1$ and $\textbf{e}^2$ orthogonal to $\vec{k}$.

Keeping the tensor perturbation at linear order and the linear scalar perturbations up to second order, one can obtain the equation of motion for each polarization $h_\lambda$ from the Einstein equation as
\begin{align}\label{eom_h}
h_\lambda^{\prime\prime}(\vec{k},\tau) + 2\mathcal{H} h_\lambda^\prime (\vec{k},\tau)+ k^2 h_\lambda(\vec{k},\tau) = 4 S_\lambda (\vec{k},\tau),
\end{align} 
where second-order perturbations are projected away in the transverse-traceless decomposition \cite{Baumann:2007zm} and we have neglected the anisotropic stress in the energy momentum tensor, so that $\Psi = \Phi$, and thus
\begin{align}
S_\lambda(\vec{k},\tau) =& \int \frac{d^3q}{(2\pi)^{3/2}} e_{ij}^\lambda(\vec{k}) q^iq^j \psi_{\vec{p}}\psi_{\vec{p}} f(p,q,\tau), \\
f(p,q,\tau) =& 2T(p\tau)T(q\tau) + \frac{4}{3(1+w)} 
\left[\frac{T^\prime(p\tau)}{\mathcal{H}} + T(p\tau)\right] \left[\frac{T^\prime(q\tau)}{\mathcal{H}} + T(q\tau)\right],
\end{align}
where $\mathcal{H} = aH = 2/ [(1+3w)\tau]$ and $w$ determines the equation of state of the fluid that fills the universe, $P=w\rho$. The two internal momenta are given by $p\equiv \vert\vec{p}\vert$ and $q\equiv \vert\vec{q}\vert$, and $\vec{k} = \vec{p} +\vec{q}$. 
The time evolution of the scalar potential is described by $\Psi_{\vec{k}}(\tau) = T(k\tau)\psi_{\vec{k}}$ with respect to the primordial value $\psi_{\vec{k}}$, where the transfer function $T(k\tau)$ in the radiation-dominated universe is given in Ref.~\cite{Domenech:2021ztg}.

The primordial Newtonian potential $\psi_{\vec{k}}$ well outside the Hubble radius is related to the (gauge-invariant) curvature perturbation 
${\cal R}$ as $\psi_{\vec{k}} = [(3 + 3 w) / (5 + 3 w) ] {\cal R}$, which yields
\begin{align}
\left\langle \psi_{\vec{k}}\psi_{\vec{K}}\right\rangle = \delta^{(3)}\left(\vec{k} +\vec{K}\right) \frac{2\pi^2}{k^3} \left(\frac{3+3w}{5+3w}\right)^2 \mathcal{P}_{\cal R}(k).
\end{align}
This is where parameters of the inflationary scenario given in \S~\ref{sec:model} enter the power spectrum of the tensor perturbation. 

Solving the equation of motion of Eq.~\eqref{eom_h} by virtue of the Green's function method of Eq.~\eqref{sol_h_Green}, we can compute the total power spectrum of the tensor perturbation as
\begin{align}\label{def_P_h}
\delta^{(3)}(\vec{k}+\vec{K}) P_h(k, \tau) 
=& \frac{k^3}{2\pi^2} \sum_\lambda^{+,\times} \left\langle h_\lambda(\vec{k},\tau)h_\lambda(\vec{K},\tau)\right\rangle, \nonumber\\
=& \frac{k^3}{2\pi^2} \int^\tau d\tau_1 G_{\vec{k}}(\tau ;\tau_1) \frac{a(\tau_1)}{a(\tau)}
 \int^\tau d\tau_2 G_{\vec{K}}(\tau;\tau_2)   \frac{a(\tau_2)}{a(\tau)} \nonumber\\
 &\qquad\times \sum_\lambda^{+,\times} \left\langle S_\lambda(\vec{k},\tau_1)S_\lambda(\vec{K},\tau_2)\right\rangle.
\end{align}
It is convenient to use the dimensionless variables $u \equiv p/k$, $v\equiv q/k$ and $z \equiv k\tau$ to rewrite the tensor spectrum as
\begin{align}\label{eq:P_h_uv_variable}
P_h(k, z) &= 4\int_{0}^{\infty} dv\int_{\vert 1-v \vert}^{1+v} du 
\left[\frac{v}{u} - \frac{(1-u^2+v^2)}{4uv}\right]^2 I^2(u,v,z) \mathcal{P}_{\cal R}(ku) \mathcal{P}_{\cal R}(kv), \\
I(u,v,z) &= \frac{9(1+w)^2}{(5+3w)^2}\int_{0}^z dz_1 \frac{a(z_1)}{a(z)} kG_{\vec{k}} (z,z_1) f(u,v,z), 
\end{align}
where our definition of $I(u,v,z)$ coincides with that defined in Ref.~\cite{Kohri:2018awv}. Note that the projection of momentum under polarization tensors can be found in the Appendix B of Ref.~\cite{Atal:2021jyo}, where 
\begin{align}
 (e_{ij}^+ q_iq_j)^2 + (e_{ij}^\times q_iq_j)^2 = k^4v^4\left[1-\frac{(1-u^2+v^2)^2}{(2v)^2}\right]^2.   
\end{align}
For numerical evaluation, we adopt new variables $t = u+v-1$, $s =u-v$ introduced in Ref.~\cite{Kohri:2018awv}, where $u=(t+s+1)/2$, $v=(t-s+1)/2$ and the tensor spectrum now reads
\begin{align}
\label{eq:powerspectrum}
P_h(k, z) =& 2\int_{0}^{\infty} dt \int_{-1}^{1} ds \left[ \frac{t (2 + t) (s^2 - 1)}{(1 - s + t) (1 + s + t)}\right]^2 \\\nonumber
&\quad \times {\cal P}_{\cal R}\left(\frac{k (t+s+1)}{2}\right) {\cal P}_{\cal R}\left(\frac{k (t-s+1)}{2}\right) I_{\rm RD}^2(s,t, z).
\end{align}
In the late-time limit of the radiation-dominated universe, that is, for $\tau\rightarrow\infty$ and $z \gg 1$, we have the oscillation-averaged result from Ref.~\cite{Kohri:2018awv}:
\begin{align}
\label{eq:GWtf}
&\overline{I^2_{\textrm{RD}}(s,t, k\tau \to \infty)} =  \frac{288 (-5 + s^2 + t (2 + t))^2}{z^2(1 - s + t)^6 (1 + s + t)^6} \times
\\
&\qquad\Bigg\{ \frac{\pi^2}{4} \left(-5 + s^2 + t (2 + t)\right)^2 \Theta\left(t-(\sqrt{3}-1)\right) +
\nonumber \\\nonumber
&\qquad \left[-(t - s + 1) (t + s + 1) + 
\frac{1}{2} (-5 + s^2 + t (2 + t)) \ln\left|\frac{(-2 + t (2 + t))}{3 - s^2}\right|\right]^2\Bigg\},
\end{align}
where $\Theta$ is the usual Heaviside theta function. Hence the averaged analytical transfer function during radiation domination in Eq.~\eqref{eq:GWtf} can be substituted into Eq.~\eqref{eq:powerspectrum}. The resulting integral will yield the oscillation-averaged power spectrum $\overline{P_h(k,\eta)}$ which can be substituted into Eq.~\eqref{eq:GWspectrum}, and the dimensionless GW background to be compared with experimental limits or signals can be determined. We have the calculation by direct comparison with the scale-invariant power spectrum normalized to unity ($A_{\cal R} = 1$), which yields a dimensionless gravitational wave spectrum of $\Omega_{\textrm{GW}}/A_{\cal R}^2 = 0.822$ as expected in Ref.~\cite{Kohri:2018awv}.

\end{widetext}


%

\end{document}